\documentclass[10pt,twocolumn,aps,prc,preprintnumbers,amsmath,amssymb,
floatfix,showpacs,superscriptaddress,bibnotes,altaffilletter,
nofootinbib]{revtex4}
\usepackage{graphicx}
\usepackage{psfig}%
\usepackage{dcolumn}
\usepackage{bm}
\usepackage{pifont}
\begin{document}

\title{Extraction of the Neutron Magnetic Form Factor from 
Quasi-elastic ${\bf ^3}\vec{\bf He}{\bf (}\vec{\bf e}{\bf ,e')}$ 
at $\bf Q^2 = 0.1 - 0.6$ (GeV/c)$\bf ^2$}

\newcommand{\caltech}{California Institute of Technology, Pasadena,
  California 91125}
\newcommand{\csla}{California State University, Los Angeles, 
  Los Angeles, California 90032}
\newcommand{\wm}{College of William and Mary, Williamsburg, Virginia 23187}
\newcommand{\duke}{Duke University, Durham, North Carolina 27708}
\newcommand{\pisa}{Istituto Nazionale di Fisica Nucleare, 
  Sezione di Pisa, I-56100 Pisa, Italy}
\newcommand{\romaone}{Istituto Nazionale di Fisica Nucleare, 
  Sezione di Roma, I-00185 Roma, Italy}
\newcommand{\romatwo}{Universit\`{a} di Roma ``Tor Vergata''
  and Istituto Nazionale di Fisica Nucleare, 
  Sezione di Roma II, I-00133 Roma, Italy}
\newcommand{\cracow}{M. Smoluchowski Institute of Physics,
  Jagellonian University, PL-30059 Cracow, Poland}
\newcommand{\kharkov}{Kharkov Institute of Physics and Technology, 
  Kharkov 310108, Ukraine}
\newcommand{\kent}{Kent State University, Kent, Ohio 44242}
\newcommand{\MIT}{Massachusetts Institute of Technology, 
  Cambridge, Massachusetts 02139}
\newcommand{\ncc}{North Carolina Central University, Durham,
  North Carolina 27707}
\newcommand{\odu}{Old Dominion University, Norfolk, Virgina 23508}
\newcommand{\princeton}{Princeton University, 
  Princeton, New Jersey 08544}
\newcommand{\bochum}{Ruhr-Universit\"{a}t Bochum, D-44780 Bochum, Germany}
\newcommand{\rutgers}{Rutgers University, Piscataway, New Jersey 08855}
\newcommand{\syr}{Syracuse University, Syracuse, New York 13244}
\newcommand{\temple}{Temple University, Philadelphia, Pennsylvania 19122}
\newcommand{\jlab}{Thomas Jefferson National Accelerator Facility, 
  Newport News, Virgina 23606}
\newcommand{\clermont}{Universit\'{e} Blaise Pascal/IN2P3, 
  F-63177 Aubi\`{e}re, France}
\newcommand{\uky}{University of Kentucky, Lexington, Kentucky 40506}
\newcommand{\unh}{University of New Hampshire, Durham, New Hampshire 03824}
\newcommand{\uva}{University of Virginia, Charlottesville, Virginia 22903}
\newcommand{\rentec}{Renaissance Technologies Inc., Stony Brook, 
  New York 11790}
\newcommand{\iucf}{Indiana University Cyclotron Facility,
  Bloomington, Indiana 47408}

\affiliation{\caltech}
\affiliation{\csla}
\affiliation{\wm}
\affiliation{\duke}
\affiliation{\pisa}
\affiliation{\romaone}
\affiliation{\romatwo}
\affiliation{\cracow}
\affiliation{\kharkov}
\affiliation{\kent}
\affiliation{\MIT}
\affiliation{\ncc}
\affiliation{\odu}
\affiliation{\princeton}
\affiliation{\bochum}
\affiliation{\rutgers}
\affiliation{\syr}
\affiliation{\temple}
\affiliation{\jlab}
\affiliation{\clermont}
\affiliation{\uky}
\affiliation{\unh}
\affiliation{\uva}

\author{B.~Anderson}    \affiliation{\kent} 
\author{L.~Auberbach}   \affiliation{\temple}
\author{T.~Averett}     \affiliation{\wm}
\author{W.~Bertozzi}    \affiliation{\MIT}
\author{T.~Black}
  \altaffiliation[Present address: ]{University of North Carolina at 
    Wilmington, Wilmington, North Carolina 28403}
  \affiliation{\MIT}
\author{J.~Calarco}     \affiliation{\unh}
\author{L.~Cardman}     \affiliation{\jlab}
\author{G.~D.~Cates}
  \altaffiliation[Present address: ]{\uva}
  \affiliation{\princeton}
\author{Z.~W.~Chai}
  \altaffiliation[Present address: ]{Brookhaven National Laboratory,
    Upton, New York 11973}
  \affiliation{\MIT}
\author{J.~P.~Chen}     \affiliation{\jlab}
\author{Seonho~Choi}
  \altaffiliation[Present address: ]{Seoul National University,
    Seoul, 151-747 Korea}
  \affiliation{\temple}
\author{E.~Chudakov}    \affiliation{\jlab}
\author{S.~Churchwell}  
  \altaffiliation[Present address: ]{University of Canterbury,
    Christchurch, New Zealand}
  \affiliation{\duke}
\author{G.~S.~Corrado}  \affiliation{\princeton}
\author{C.~Crawford}    \affiliation{\MIT}
\author{D.~Dale}        \affiliation{\uky}
\author{A.~Deur}        
  \altaffiliation[Present address: ]{\jlab}
  \affiliation{\clermont}
\author{P.~Djawotho}
  \altaffiliation[Present address: ]{\iucf}
  \affiliation{\wm}
\author{D.~Dutta}
  \altaffiliation[Present address: ]{\duke}
  \affiliation{\MIT}
\author{J.~M.~Finn}     \affiliation{\wm} 
\author{H.~Gao}
  \altaffiliation[Present address: ]{\duke}
  \affiliation{\MIT} 
\author{R.~Gilman}      \affiliation{\rutgers} \affiliation{\jlab}
\author{A.~V.~Glamazdin}\affiliation{\kharkov} 
\author{C.~Glashausser} \affiliation{\rutgers}
\author{W.~Gl\"{o}ckle} \affiliation{\bochum}
\author{J.~Golak}       \affiliation{\cracow}
\author{J.~Gomez}       \affiliation{\jlab}
\author{V.~G.~Gorbenko} \affiliation{\kharkov}
\author{J.-O.~Hansen}   
  \email[Corresponding author. Electronic mail address: ]{ole@jlab.org}
  \affiliation{\jlab}
\author{F.~W.~Hersman}  \affiliation{\unh}
\author{D.~W.~Higinbotham}
  \altaffiliation[Present address: ]{\jlab}
  \affiliation{\uva}
\author{R.~Holmes}      \affiliation{\syr} 
\author{C.~R.~Howell}   \affiliation{\duke}
\author{E.~Hughes}      \affiliation{\caltech}
\author{B.~Humensky}
  \altaffiliation[Present address: ]{The Enrico Fermi Institute,
    University of Chicago, Chicago, Illinois 60637}
  \affiliation{\princeton}
\author{S.~Incerti}
  \altaffiliation[Present address: ]{IN2P3/CNRS, Universit\'{e} Bordeaux,
    F-33175 Gradignan Cedex, France}
  \affiliation{\temple} 
\author{C.~W.~de Jager}  \affiliation{\jlab}
\author{J.~S.~Jensen}   
  \altaffiliation[Present address: ]{Bookham Inc., San Jose, California 95134}
  \affiliation{\caltech}
\author{X.~Jiang}       \affiliation{\rutgers} 
\author{C.~E.~Jones}
  \altaffiliation[Present address: ]{Siimpel Corp., Arcadia,
    California 91006}
  \affiliation{\caltech}
\author{M.~Jones}
  \altaffiliation[Present address: ]{\jlab}
  \affiliation{\wm}
\author{R.~Kahl} 
  \affiliation{\syr}
\author{H.~Kamada}
  \altaffiliation[Present address: ]{Kyushu Institute of Technology,
    Ki\-ta\-ky\-u\-shu 804-8550, Japan}
  \affiliation{\bochum} 
\author{A.~Kievsky}     \affiliation{\pisa} 
\author{I.~Kominis}
  \affiliation{\princeton}
\author{W.~Korsch}      \affiliation{\uky}
\author{K.~Kramer}
  \altaffiliation[Present address: ]{\duke}
  \affiliation{\wm}
\author{G.~Kumbartzki}  \affiliation{\rutgers}
\author{M.~Kuss}
  \altaffiliation[Present address: ]{\pisa}
  \affiliation{\jlab} 
\author{E.~Lakuriqi}
  \altaffiliation[Present address: ]{University of Pennsylvania, Philadelphia,
  Pennsylvania 19104}
  \affiliation{\temple}
\author{M.~Liang}
  \affiliation{\jlab}
\author{N.~Liyanage}
  \altaffiliation[Present address: ]{\uva}
  \affiliation{\jlab}
\author{J.~LeRose}      \affiliation{\jlab} 
\author{S.~Malov}
  \affiliation{\rutgers}
\author{D.~J.~Margaziotis} \affiliation{\csla}
\author{J.~W.~Martin}
  \altaffiliation[Present address: ]{University of Winnipeg,
    Winnipeg, Manitoba R3B 2E9, Canada}
  \affiliation{\MIT}
\author{K.~McCormick}
  \altaffiliation[Present address: ]{Pacific Northwest Laboratory,
    Richland, Washington 99352}
  \affiliation{\odu}
\author{R.~D.~McKeown}  \affiliation{\caltech}
\author{K.~McIlhany}
  \altaffiliation[Present address: ]{United States Naval Academy, 
    Annapolis, Maryland 21402}
  \affiliation{\MIT}
\author{Z.-E.~Meziani}  \affiliation{\temple}
\author{R.~Michaels}    \affiliation{\jlab}
\author{G.~W.~Miller}   \affiliation{\princeton}
\author{J.~Mitchell}
  \altaffiliation[Present address: ]{\rentec}
  \affiliation{\jlab}
\author{S.~Nanda}       \affiliation{\jlab} 
\author{E.~Pace}        \affiliation{\romatwo}
\author{T.~Pavlin}
  \altaffiliation[Present address: ]{Center for Astrophysics, 
      Harvard University, Cambridge, Massachusetts 02138}
  \affiliation{\caltech}
\author{G.~G.~Petratos} \affiliation{\kent} 
\author{R.~I.~Pomatsalyuk} \affiliation{\kharkov}
\author{D.~Pripstein}
  \altaffiliation[Present address: ]{Harmonic Inc., Sunnyvale, 
    California 94089}
  \affiliation{\caltech}
\author{D.~Prout}
  \affiliation{\kent}
\author{R.~D.~Ransome}  \affiliation{\rutgers}
\author{Y.~Roblin}
  \altaffiliation[Present address: ]{\jlab}
  \affiliation{\clermont} 
\author{M.~Rvachev}
  \affiliation{\MIT}
\author{A.~Saha}        \affiliation{\jlab}
\author{G.~Salm\`{e}}   \affiliation{\romaone}
\author{M.~Schnee}
  \affiliation{\temple} 
\author{J.~Seely}       \affiliation{\MIT}
\author{T.~Shin}
  \altaffiliation[Present address: ]{Hampton University, Hampton,
  Virginia 23668}
  \affiliation{\MIT} 
\author{K.~Slifer}
  \altaffiliation[Present address: ]{\uva}
  \affiliation{\temple}
\author{P.~A.~Souder}   \affiliation{\syr}
\author{S.~Strauch}
  \altaffiliation[Present address: ]{The George Washington University, 
    Washington, District of Columbia 20052}
  \affiliation{\rutgers}
\author{R.~Suleiman}
  \altaffiliation[Present address: ]{Virginia Polytechnic Institute, 
    Blacksburg, Virginia 24061}
  \affiliation{\kent} 
\author{M.~Sutter}
  \altaffiliation[Present address: ]{RightAnswers LLC, Clark,
    New Jersey 07066}
  \affiliation{\MIT}
\author{B.~Tipton}
  \altaffiliation[Present address: ]{MIT-Lincoln Laboratory, Lexington,
    Massachusetts 02420}
  \affiliation{\MIT}
\author{L.~Todor}
  \altaffiliation[Present address: ]{University of Richmond, Richmond, 
  Virginia 23173}
  \affiliation{\odu}
\author{M.~Viviani}     \affiliation{\pisa} 
\author{B.~Vlahovic}    \affiliation{\ncc} \affiliation{\jlab}
\author{J.~Watson}      \affiliation{\kent}
\author{C.~F.~Williamson} \affiliation{\MIT}
\author{H.~Wita{\l}a}   \affiliation{\cracow} 
\author{B.~Wojtsekhowski} \affiliation{\jlab}
\author{F.~Xiong}
  \altaffiliation[Present address: ]{\rentec}
  \affiliation{\MIT}
\author{W.~Xu}
  \altaffiliation[Present address: ]{\duke}
  \affiliation{\MIT}
\author{J.~Yeh}
  \affiliation{\syr} 
\author{P.~\.{Z}o{\l}nierczuk} 
  \altaffiliation[Present address: ]{\iucf}
  \affiliation{\uky}
\collaboration{Jefferson Lab E95-001 Collaboration} \noaffiliation

\date{20 November 2006}

\begin{abstract}

We have measured the transverse asymmetry $A_{T'}$
in the quasi-elastic $^3\vec{\rm He}(\vec{e},e')$ process
with high precision at $Q^2$-values from 0.1 to 0.6 (GeV/c)$^2$. 
The neutron magnetic form factor
$G_M^n$ was extracted at $Q^2$-values of 0.1 and 0.2 (GeV/c)$^2$ using a 
non-relativistic Faddeev calculation which includes both final-state 
interactions (FSI) and 
meson-exchange currents (MEC). Theoretical uncertainties due to
the FSI and MEC effects were constrained with a precision measurement 
of the spin-dependent asymmetry in the threshold region of 
$^3\vec{\rm He}(\vec{e},e')$. We also extracted the neutron magnetic form 
factor $G_M^n$ at $Q^2$-values of 0.3 to 0.6 (GeV/c)$^2$ based on
Plane Wave Impulse Approximation calculations. 

\end{abstract}

\maketitle
\section{introduction}
The electromagnetic structure of the nucleon has long been a topic of 
fundamental interest in nuclear and particle physics.
First-order nucleon electromagnetic properties
are commonly parameterized in terms of elastic form factors \cite{gao_review}.
At low values of four-momentum transfer squared, $Q^2$, these functions
have a simple interpretation as the Fourier transforms of the
nucleon charge and magnetization densities in the Breit frame.
Their precise experimental determination is important both for testing 
fundamental theories 
of hadron structure and for the analysis of other experiments in the field,
such as parity violation measurements~\cite{sample,happex} that are 
designed to probe the strangeness content of the nucleon.

The proton form factors have been determined with
good precision at low $Q^2$ using Rosenbluth separation of elastic
electron-proton cross sections, and more recently at higher $Q^2$
using a polarization transfer technique~\cite{jones,gayou}.
The neutron form factors are known less well
because of the zero electric charge of the neutron,
causing its electric form factor to be small, and
experimental complications such as the 
lack of free neutron targets and difficulties associated with
neutron detectors.

Over the past two decades, with the advent of much improved experimental 
facilities, the precise measurement of both the
neutron electric form factor, $G_E^n$, and the magnetic form factor,
$G_M^n$, has become a focus of activity. 
Until recently, most data on $G_M^n$ had been deduced from elastic and
quasi-elastic electron-deuteron scattering.  Inclusive
measurements of this type suffer from large theoretical uncertainties
due in part to the deuteron model employed and in part to 
corrections for final-state
interactions (FSI) and meson-exchange currents (MEC).  The sensitivity
to nuclear structure is reduced by measuring the neutron in coincidence,
$^2$H$(e,e'n)$ \cite{Mark93}, and, further, 
by taking the ratio of cross sections of $^2$H$(e,e'n)$ to $^2$H$(e,e'p)$
at quasi-elastic kinematics \cite{Ankl94,Brui95,Schoch2000,Ankl98,Kubon2002}.  
Uncertainties of less than 2\% in $G_M^n$ have been achieved
in the region $Q^2 < 1$~(GeV/c)$^2$ using the latter technique
\cite{Ankl98,Kubon2002}.  Despite this high precision, 
there is significant disagreement between the 
results of \cite{Mark93,Brui95, Schoch2000} and those
of the more recent experiments \cite{Ankl94,Ankl98,Kubon2002} 
of up to 10\% in the absolute value of $G_M^n$. An
explanation has been suggested in \cite{jourdan}, but the issue
has remained contentious.

To clarify the situation experimentally, additional data on $G_M^n$, 
preferably obtained using a complementary method, are highly desirable.
Inclusive quasi-elastic $^3\vec{\rm He}(\vec{e},e')$ scattering
provides such an alternative approach \cite{Blankleider84}.  
In contrast to
deuterium experiments, this technique employs a different target and
relies on polarization degrees of freedom.  It is thus subject to
completely different systematics.
On the other hand, due to the more complex physics of the three-body system,
the precise extraction of nucleon form factors from 
polarized $^3$He measurements requires careful modeling of the 
nuclear structure and of the reaction mechanism.
Recent advances in Faddeev calculations 
\cite{Faddeev61,Golak01,golak2} have brought
theoretical uncertainties of $^3$He models sufficiently 
under control to allow such studies
in the non-relativistic kinematic regime. 
A precision comparable to that of the deuterium ratio experiments 
can be achieved using the polarized $^3$He technique \cite{Xu2000}.

The use of polarized $^3$He targets was pioneered
at MIT-Bates \cite{Gao94,Jones93,Thompson92,Hansen95} and
Mainz \cite{Meyerhoff94}. 
In \cite{Gao94}, $G_M^n$ was extracted for the first time
from quasi-elastic inclusive scattering from polarized $^3$He,
although with a large statistical uncertainty.

In this paper, we report on the first precision measurement 
of the so-called transverse asymmetry $A_{T'}$, which is sensitive to
$G_M^n$, in the inclusive reaction 
$^3\vec{\rm He}(\vec{e},e')$. The results were obtained in
Hall A at the Thomas Jefferson National Accelerator Facility
(Jefferson Lab). Brief reports of these data have
appeared previously \cite{Xu2000,Xu2003,xiong}. This paper
presents the data analysis and evaluation of model uncertainties
in much more detail. In addition, the analysis has been slightly refined.
The results presented here are final.

The neutron magnetic form factor $G_M^n$ was extracted at
$Q^2 = 0.1$ to $0.6$ (GeV/c)$^2$ in steps of 0.1 (GeV/c)$^2$
\cite{Xu2000,Xu2003}. In addition, high-precision asymmetry 
data in the $^3$He breakup region were obtained at $Q^2$-values of 
0.1 and 0.2 (GeV/c)$^2$ \cite{xiong}.
The threshold data provide a stringent test of the above-mentioned Faddeev 
calculations because they cover a kinematical region where the proper 
treatment of the reaction mechanism is particularly important. 

At the $Q^2 = 0.1$ and $0.2$ (GeV/c)$^2$ kinematics, 
$G_M^n$ was extracted using a state-of-the-art Faddeev 
calculation \cite{Golak01}. 
At these low $Q^2$, relativistic effects are small, and the 
non-relativistic Faddeev results have been shown to be in good
agreement with a diverse set of few-body data, including our own
$^3$He breakup threshold data \cite{xiong}.
On the other hand, the extraction of $G^n_M$ from our
$^3$He asymmetry data at higher values of $Q^2$ with the same precision
as that achieved at low $Q^2$ would require a more advanced theory 
that includes both an accurate treatment of reaction mechanism effects
(FSI and MEC) and proper relativistic corrections (and possibly
other refinements, such as $\Delta$-isobar excitations, presumed to be 
small at our kinematics).
Unfortunately, such a comprehensive calculation is not available at the 
present time, and efforts to extend the theory are only in the
beginning stages. For example, full inclusion of FSI 
has been investigated for the two-body channel in \cite{Kievsky04}.
The Hannover group has carried out a coupled-channel calculation
of $^3\vec{\rm He}(\vec{e},e')$ that accounts for FSI and $\Delta$-isobars
\cite{SauerFSI04}, unfortunately also with limited success 
at higher $Q^2$. Nonetheless, we observe that the size of
FSI and MEC corrections to inclusive scattering data near the top 
of the quasi-elastic peak has been predicted to 
diminish sharply with increasing momentum transfer 
\cite{JRA,Pace91,Benhar99,Arenhovel93}. Hence, it appears likely that the
Plane Wave Impulse Approximation (PWIA), in which the knocked-out 
nucleon is described by a plane wave while the spectator pair is 
fully interacting,
is reasonably accurate at the higher $Q^2$-values of this experiment.
A quantitative estimate of the $Q^2$-behavior
of deviations from the PWIA, in particular of the size of FSI corrections, 
could be obtained by performing a $y$-scaling analysis on the 
present $^3$He asymmetry data \cite{Ciofi91}. Such an analysis may be
carried out in a future publication.

Taking the pragmatic point of view that the PWIA is currently the best
available theory describing inclusive quasi-elastic scattering from polarized
$^3$He at $Q^2 \geq 0.3$~(GeV/c)$^2$, we have extracted $G^n_M$ from our
higher $Q^2$ data \cite{Xu2003} using PWIA.
While we do not attempt to go beyond the PWIA by
computing corrections for the various effects omitted in this approximation, 
we provide estimates of the uncertainties of the results 
in considerable detail.
Despite the relatively large theoretical uncertainties in this approach,
our results are in good agreement with the recent deuterium
ratio measurements from Mainz \cite{Ankl98,Kubon2002} in the same
$Q^2$-region. 

\section{Theory}
\label{sec:theory}
     
\subsection{Spin-dependent Inclusive Electron Scattering}

\begin{figure}[bt]
\begin{center}
\includegraphics[width=0.45\textwidth]{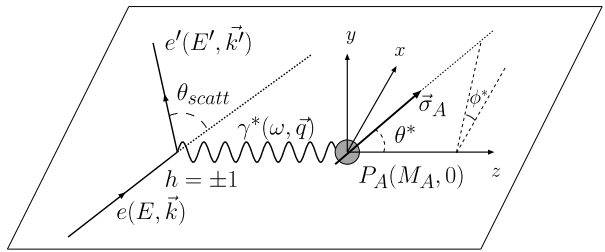}
\end{center}
\caption{\small{Spin-dependent inclusive electron scattering from a 
polarized target. The target spin angles, $\theta^\ast$ and $\phi^\ast$
are defined with respect to the three-momentum transfer vector $\bf q$}.}
\label{fig:scatt}
\end{figure}

Figure \ref{fig:scatt} depicts inclusive scattering of longitudinally 
polarized electrons from a polarized nuclear target. 
The four-momentum of the electrons before and after the reaction is
$k = (E, {\bf k})$ and $k' = (E', {\bf k'})$, respectively.
The four-momentum transfer to the target is
$q = k-k' = (\omega,{\bf q})$, with the usual definition
$Q^2 \equiv -q^2$. 

The experiment measured the spin-dependent asymmetry
$A = (\sigma^+-\sigma^-)/(\sigma^++\sigma^-)$,
where $\sigma^\pm$ is the differential cross section for quasi-elastic
scattering of electrons with helicity $h = \pm 1$ from polarized $^3$He.
It can be expressed in terms of nuclear response functions, $R(Q^2,\omega)$,
and kinematic factors, $v(Q^2,\omega)$, as \cite{donnelly1}
\begin{equation}
A = -\frac{\cos{\theta^\ast}\nu_{T'}R_{T'} +
  2\sin{\theta^\ast}\cos{\phi^\ast}\nu_{\it TL'}R_{\it TL'}}{\nu_{L}R_{L} +
  \nu_{T}R_{T}},
\label{eq:asym}
\end{equation}  
where $\theta^\ast$ and $\phi^\ast$ are the polar and azimuthal angles of the 
target spin direction with respect to the three-momentum transfer vector, 
{$\bf q$}, as shown in Figure~\ref{fig:scatt}.
By choosing $\theta^\ast$ = $0^{\circ}$ or $\theta^\ast = 90^{\circ}$, 
one can select the 
transverse asymmetry, $A_{T'}$, or the longitudinal-transverse asymmetry,
 $A_{\it TL'}$. 

The nuclear response functions for inclusive quasi-el\-as\-tic 
scattering have been
obtained through both PWIA and Faddeev calculations. These 
calculations will be discussed briefly next.

\subsection{Plane Wave Impulse Approximation}
\label{sec:pwia}

In the PWIA, it is assumed that a single
nucleon within the target nucleus completely absorbs the 
momentum of the virtual photon and leaves the interaction region
as a plane wave. The remaining two-nucleon subsystem still
undergoes interaction. Exchange current effects are ignored.
The target nucleus, in our case $^3$He, however, is described
by the solution of the Schr\"{o}dinger equation with realistic
nuclear forces. Relativistic effects are included by using
relativistic energy conservation and a relativistic electron-nucleon
cross section.  

The nuclear current tensor is calculated
as the product of the nucleonic current tensor and the nuclear spectral 
function, which contains the nuclear structure information
(see for example \cite{Kievsky97,Ciofi93,Ciofi95,schulze1}).
The spin-independent part of the spectral function has the well-known
interpretation as the probability of finding a nucleon of certain
momentum and isospin in the target nucleus \cite{hajduk1}.
The PWIA formalism available in the literature 
is largely but not necessarily fully covariant.

Expressions for the matrix elements of the nucleonic current tensor 
and the spin-dependent nuclear spectral function
have been derived in \cite{schulze1}. The spectral function 
can be computed numerically from the nuclear wave function, which in turn
can be obtained from a model of the nucleon-nucleon (NN) potential. 
With the nucleonic current tensor and the nuclear spectral function at hand, 
expressions for the functions $R(Q^2,\omega)$, required
in (\ref{eq:asym}), can be derived and evaluated numerically
\cite{schulze1}. 

The PWIA results presented in the paper were calculated following
\cite{Kievsky97}. The calculation was based on a $^3$He
wave function derived from the Argonne AV18 NN potential \cite{wiringa1}
and used the H\"{o}hler nucleon form factor parameterization \cite{hoh}.

\subsection{Non-relativistic Faddeev Calculation}

\begin{figure}
\includegraphics[height=0.235\textheight]{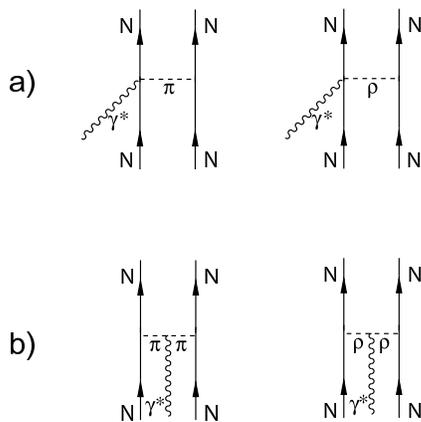}
\caption{\small{Meson-exchange current contributions included in
the Faddeev calculation \protect\cite{Golak01}. a) Couplings to a
correlated nucleon pair; b) couplings to a $\pi$ or $\rho$ in flight.}}
\label{fig:mec}
\end{figure}

In the Faddeev approach \cite{Faddeev61}, the coordinate-space 
Schr{\"o}dinger equation for three nucleons with two-nucleon interactions 
is decomposed into three separate equations \cite{carlson1}. 
In momentum space, the three Faddeev equations
can be written as three integral equations. 
The kernel in each equation involves only 
the interaction between one pair of the nucleons. 
Solutions are obtained numerically.
The Faddeev decomposition of the three-body (and four-body) problem has proven
to be a very useful computational tool in studies of light nuclei. 

With regard to $^3$He, the Faddeev formalism has been applied to 
unpolarized $pd$ and $ppn$ electrodisintegration 
\cite{ishikawa1,golak2} with full inclusion of all
final-state rescattering processes.
This calculation was subsequently 
extended to electrodisintegration of polarized $^3$He \cite{ishikawa2}.
A further extension was made by including proper treatment of
meson-exchange currents \cite{Golak01} 
according to the Riska prescription \cite{riska1}, which
relates NN forces and meson-exchange currents in a 
model-independent manner through the continuity equation. 
In \cite{Golak01}, only the dominant $\pi$- and $\rho$-like meson 
exchange terms shown in Figure~\ref{fig:mec} were considered.
The effect of $\Delta$ currents has also been studied
and found to be small (see Section~\ref{sec:thresholdasym}).

The derivation of the nuclear response functions
in the Faddeev approach is described further in \cite{ishikawa2}.
In this work, the resulting expressions were evaluated numerically using the
framework of \cite{Golak01} for a large number of kinematical points
corresponding to the acceptance regions covered by the experiment.
The underlying $^3$He wave function was obtained using the BonnB NN 
potential \cite{bonnb}. Again, the H\"{o}hler parameterization 
\cite{hoh} was used to model the nucleon elastic form factors.
The Faddeev calculation does not include relativistic effects.

\subsection{Extraction of the Neutron Magnetic Form Factor}

Because the $^3$He nuclear spin is carried mainly by the neutron,
the spin-dependent response functions $R_{T'}$ and $R_{\it TL'}$ can
be expected to contain a large if not dominant neutron contribution
at quasi-elastic kinematics \cite{Blankleider84}.
Comparison of Equation~(\ref{eq:asym}) with the corresponding
expression for scattering from a free nucleon leads to the expectation
(within PWIA) that
\begin{eqnarray}
R_{T'}      &\propto& P_n (G_M^n)^2 + P_p (G_M^p)^2 
   \label{eq:rt_propto} \\
R_{\it TL'} &\propto& P_n G_M^n G_E^n + P_p G_M^p G_E^p ,
   \label{eq:rtl_propto}
\end{eqnarray}
where $P_n$  and $P_p$ are the effective polarizations of 
the neutron and the protons, respectively, in $^3$He. Because the proton spins
largely cancel, we have $|P_p| \ll |P_n|$. Effective polarizations
have been calculated {\it e.g.}\/ in \cite{Friar90,Kievsky97,Ciofi93}.
Since $|G_M^n| \approx |G_M^p|$, the proton 
contribution to the transverse response $R_{T'}$ is small, and
hence $R_{T'}$ is essentially proportional to $(G_M^n)^{2}$.
Based on these arguments, the asymmetry
$A_{T'}$ defined in (\ref{eq:asym}) can be written as a 
function of the neutron magnetic form factor,
\begin{equation}
        A_{T'} (G_M^n {^{2}}) =\frac{1+a(G_M^n)^2}{b+c(G_M^n)^{2}},
\label{eq:at_heuristic}
\end{equation}
where $|a|\gg 1$ and $b > c$ at low $Q^2$ where the above assumptions
hold. By comparing $A_{T'}$ data 
with predictions for $A_{T'}$ from a calculation, one can extract $G_M^n$.
The detailed procedure will be discussed in Section~\ref{sec:ffextraction}.

For completeness we mention that, because 
$|G_E^p| \gg |G_E^n|$, the proton contribution
to the transverse-long\-i\-tu\-di\-nal response, $R_{\it TL'}$, 
may be dominant despite the small effective proton polarization.
Thus inclusive scattering from polarized $^3$He is not a promising
technique to measure the neutron electric form factor, $G_E^n$
\cite{Hansen95,Golak2002}.

\section{Experiment}

\subsection{Overview \& Kinematics}

The experiment, E95-001, was performed in Hall A at Jefferson Lab
using a continuous-wave electron beam of 15~$\mu$A current and
70\% longitudinal polarization, incident on a high-pressure polarized
$^3$He gas target.  The beam energies were 778 and 1727 MeV.

Electrons scattered from the target were detected by two 
high-resolution spectrometers (HRS) positioned on the left and
right-hand side of the beam line, respectively.
Both spectrometers were configured for electron detection and for
independent operation (single-arm mode).  The ``electron
spectrometer'' on the left side of the beam performed the main physics
measurement of inclusive $^3\vec{\rm He}(\vec{e},e')$ 
scattering at six different quasi-elastic kinematics.
The second HRS, the ``hadron spectrometer'' to the right of
the beam, detected $^3\vec{\rm He}(\vec{e},e)$ elastic scattering and
provided continuous high-precision monitoring of beam and target
polarizations.  The kinematic settings are listed in
Table~\ref{tab:kinematics}.

\begin{table}[b]
\begin{center}
{Electron arm (quasi-elastic)}
\end{center}
\begin{center}
\begin{tabular}{|cccc|}
\hline\hline
$Q^2$   &  $E$  &  $E'$  &  $\theta$  \\
(GeV/c)$^2$  & (GeV) &  (GeV)  & (deg) \\
\hline
0.1  &  0.778  &  0.717  &  24.44    \\
0.193  &  0.778  &  0.667  &  35.50  \\
0.3  &  1.727  &  1.559  &  19.21    \\
0.4  &  1.727  &  1.506  &  22.62    \\
0.5  &  1.727  &  1.453  &  25.80    \\
0.6  &  1.727  &  1.399  &  28.85    \\
\hline\hline
\end{tabular}
\end{center}
\begin{center}
{Hadron arm (elastic)}
\end{center}
\begin{center}
\begin{tabular}{|cccc|}
\hline\hline
$Q^2$& $E$  &  $E'$  &  $\theta$ \\
(GeV/c)$^2$ & (GeV)  &  (GeV)  &  (deg)  \\
\hline
0.1 & 0.778 & 0.760 & 23.73 \\
 
0.2 & 1.727 & 1.691 & 15.04 \\
\hline\hline
\end{tabular}
\end{center}
\caption{\small{Kinematic settings for the quasi-elastic and
elastic measurements.}}
\label{tab:kinematics}
\end{table}

\subsection{Polarized Electron Source \& Beam Line}

The electron beam originated from a 
laser-driven ``strained'' GaAs source \cite{Alley95,Prepost95}.
Polarized electrons were produced by illuminating a GaAs crystal
in ultra-high vacuum with high-intensity circularly 
polarized laser light and removing electrons excited within the 
crystal by a strong external electric field. 
The polarization of the laser light
was controlled electronically with the help of a Pockels cell. 
In this way, the electron beam helicity could be reversed rapidly
(typically at 30 Hz), minimizing systematic errors in the measurement of
spin-dependent asymmetries.
To reduce systematic errors further, the overall sign of the beam helicity
was reversed periodically by inserting a half-wave plate into the injector
laser light path. 

The standard Hall A beam line instrumentation and beam raster \cite{halla_NIM}
was employed. The beam energy was determined with an accuracy of
better than 0.1\% for all kinematics.

\subsection{Polarized $^3$He Target}
\label{sec:poltarget}

The experiment employed an optically-pumped polarized $^3$He gas target 
\cite{halla_NIM} of the spin-exchange type \cite{Bouchiat60}.
The target cell of this system contained high-pressure ($\approx$ 10 atm) 
$^3$He gas as well as admixtures of rubidium (to facilitate
optical pumping) and nitrogen (to quench radiation trapping).
While background from the rubidium was negligible,
the nitrogen admixture contributed on the order of $10^{-2}$ to the total 
target number density, requiring a small dilution correction (see
Section~\ref{sec:dilution}). 

The target cell proper was a 40~cm long aluminum-silicate glass cell
($\rho = 2.76\ \mbox{g/cm}^3$) with $\approx 1.2$~mm thick walls
and $\approx 135~\mu$m thick end windows.
A second target cell, the so-called reference cell, was available for
calibration measurements. The reference cell had essentially 
the same dimensions as the target cell, except that it had no
thin end windows but rather a uniform glass thickness throughout.
Further details can be found in \cite{Steffen2000,Ioannis2001}.

A typical $^3$He nuclear polarization of 40\% was a\-chieved.
The target spin direction was
either $-62.5^\circ \pm 0.5^\circ$ or $-243.6^\circ \pm 0.5^\circ$ 
in the laboratory. (The difference of the 
two angles was not exactly 180$^\circ$ because of a calibration inaccuracy.)
The target spin was reversed regularly throughout the 
experiment to reduce systematic errors from false asymmetries.

\subsection{Spectrometers}
\label{sec:spectro}

The two spectrometers were equipped with their standard detector packages
\cite{halla_NIM} consisting of a pair of Vertical Drift
Chambers (VDCs) for tracking, two segmented scintillator planes to 
generate the trigger and provide time-of-flight
information, and a CO$_2$ gas Cherenkov detector for electron/pion
separation.  The HRSs had a usable momentum acceptance of approximately 9\%.
For further pion rejection, a preshower and a total-absorption shower
counter were employed in the electron-arm HRS, while the hadron-arm HRS was 
instrumented with two thin lead-glass shower counters.
The geometric solid angle of each HRS was limited to 6.0~msr by a
rectangular tungsten collimator.  The central scattering angle was
surveyed to better than 0.1~mrad.

Trajectories of scattered particles were reconstructed
using the VDC data and the standard optics model of the HRS \cite{halla_NIM}.
The achieved momentum and scattering angle resolutions ($\sigma$) were better 
than 0.05\% and 2~mrad, respectively. The transverse ({\it i.e.}\ 
along the beam) position resolution at the target was approximately 2~mm. 

The pion rejection factor with the Cherenkov detectors alone was of order 100.
Combining the Cherenkov and shower counters, a factor of over 1000 was
achieved. Pion rejection was only a concern with the left-arm HRS,
where pion production was not kinematically suppressed.


\section{Analysis}
\label{sec:analysis}

\subsection{Overview}
The experimental raw asymmetry was calculated as
\begin{equation}
        A^{exp} = \frac{N_+ - N_-}{N_+ + N_-} 
\label{eq:rawasym}
\end{equation}
where $N_+$ and $N_-$ are the electron yields normalized by charge and 
electronic live time for positive and negative electron helicities, 
respectively. 

To extract the physics asymmetry, corrections had to be made for
dilution, background, radiative effects, and bin centering.  Sources
of dilution were the finite beam and target polarizations, and
scattering from the target walls and from the nitrogen gas in the
target.  Polarized background arose from the elastic radiative tail,
which extended into the quasi-elastic region. Radiative corrections
had to be applied to the raw quasi-elastic asymmetry. Bin centering
corrections account for finite experimental acceptances.

The normalized yields in (\ref{eq:rawasym}) can be written as
\begin{equation}
N = N^{qe} + N^{ert} + N^{emp} + N^{N_2}, 
\label{eq:yields}
\end{equation}
where $N^{qe}$, $N^{ert}$, $N^{emp}$, and $N^{N_2}$ are the
contributions of quasi-elastic scattering from $^3$He (before
radiative and bin centering corrections), the elastic radiative tail,
target wall (``empty target'') scattering, and scattering from
nitrogen in the target cell, respectively. Using (\ref{eq:yields}),
one can define dilution factors for each of the three background
contributions,
\begin{eqnarray}
R^{emp} &=& \frac{N^{emp}}{N^{qe}+N^{ert}}, \label{eq:remp} \\
R^{N_2} &=& \frac{N^{N_2}}{N^{qe}+N^{ert}}, \label{eq:rn2} \\
R^{ert} &=& \frac{N^{ert}}{N^{qe}}, \label{eq:ert}
\end{eqnarray}
and express the physics asymmetry as
\begin{eqnarray}
 A^{phys} &=&~~(1+R^{ert})(1+R^{emp}+R^{N_2})\frac{A^{exp}}{P_bP_t} \nonumber\\
   & & - R^{ert} A^{ert} +\Delta A^{qe} + \Delta A^{bin},
\label{eqn_asy}
\end{eqnarray}
where $P_bP_t$ is the product of beam and target polarizations,
$A^{ert}$ is the asymmetry of the elastic radiative tail, 
$\Delta A^{qe}$ is the radiative correction to the quasi-elastic 
asymmetry, and $\Delta A^{bin}$, the bin centering correction.
In Equation (\ref{eqn_asy}), it is assumed that both the empty target
and the $N_2$ contributions have no asymmetry.  During the analysis,
the empty target and N$_2$ false asymmetries were verified to be indeed
consistent with zero.

Among the various factors in (\ref{eqn_asy}), $A^{exp}$, $R^{emp}$, and
$R^{N_2}$ could be determined directly from data, while $R^{ert}$, 
$A^{ert}$, $\Delta A^{qe}$ and $\Delta A^{bin}$ had to be determined from
calculations or simulations. $P_bP_t$ was monitored continuously during 
the experiment via elastic polarimetry and was determined as the ratio between 
the measured elastic asymmetry and the simulated elastic asymmetry,
as described in Section~\ref{sec:elastic_polarimetry}.

\subsection{Raw Asymmetries}
\label{sec:rawasym}
Raw asymmetries for both spectrometers were calculated according to 
Equation~(\ref{eq:rawasym}). 
The quasi-elastic data were analyzed in terms of electron energy loss,
$\omega = E-E'$, and grouped in bins of 10, 20, or 18.75 MeV width, depending
on $Q^2$. 
The elastic data from the right-arm spectrometer
were analyzed in terms of excitation energy, defined as
\begin{equation}
E_x = \sqrt{M^2 + 2M(E-E') - 4 EE' \sin^2(\theta/2)} - M,
\label{eq:ex}
\end{equation}
where $M$ is the mass of the $^3$He nucleus and 
$\theta$ the measured electron scattering angle. The raw elastic asymmetry
was obtained from the region $-1~\mbox{MeV} \leq E_x \leq +1~\mbox{MeV}$.

The angle between momentum transfer and
target spin, $\theta^\ast$ in Equation~(\ref{eq:asym}), varied between
0.2$^\circ$ and 10.0$^\circ$ depending on $Q^2$.  This resulted in an
$R_{TL'}$ contribution to the experimental asymmetry of less than 2\%, 
as estimated by a PWIA calculation. The $R_{TL'}$
contribution is included in the theoretical calculations that were used
to extract $G_M^n$. Even though theoretical predictions of
$R_{TL'}$ are less accurate than those of $R_{T'}$
(because of the uncertainty in $G_E^n$), the uncertainty in our extracted 
$G_M^n$ due to $R_{TL'}$ is negligible.

Raw asymmetries obtained for the four different combinations of of
target spin orientation and overall beam helicity sign were
compared to check for false asymmetries. No statistically significant false
signal was found. For the main physics 
analysis, data from the four polarization configurations were combined
to minimize the statistical uncertainty.

\subsection{Empty Target and Nitrogen Dilution Factors}
\label{sec:dilution}
Because the target cell was sealed, background from the target cell
wall could not be measured directly by emptying the target.  In
addition, the background rate from the nitrogen buffer gas in the target
could not be easily calculated because the nitrogen partial pressure 
could only be determined approximately when the cell was filled.
Therefore, it was necessary to determine both background yields in 
separate calibration runs with the reference cell. 

For each kinematics, quasi-elastic data were taken with the reference 
cell empty and filled with N$_2$ at several pressure values.
The reference cell nitrogen yield as a function of nitrogen pressure 
was determined by subtracting the
empty cell yield from the raw yields of the nitrogen runs.
As the reference cell had physical dimensions very similar to those of the
target cell, the reference cell nitrogen spectra could be used as a direct
measure of the target cell nitrogen yield, $N^{N_2}$, provided 
that they were scaled to the nitrogen pressure inside the target cell. 

\begin{figure}[btp]
\begin{center}
{$\bf N_2$}
\end{center}
\begin{center}
\includegraphics[width=0.4\textwidth]{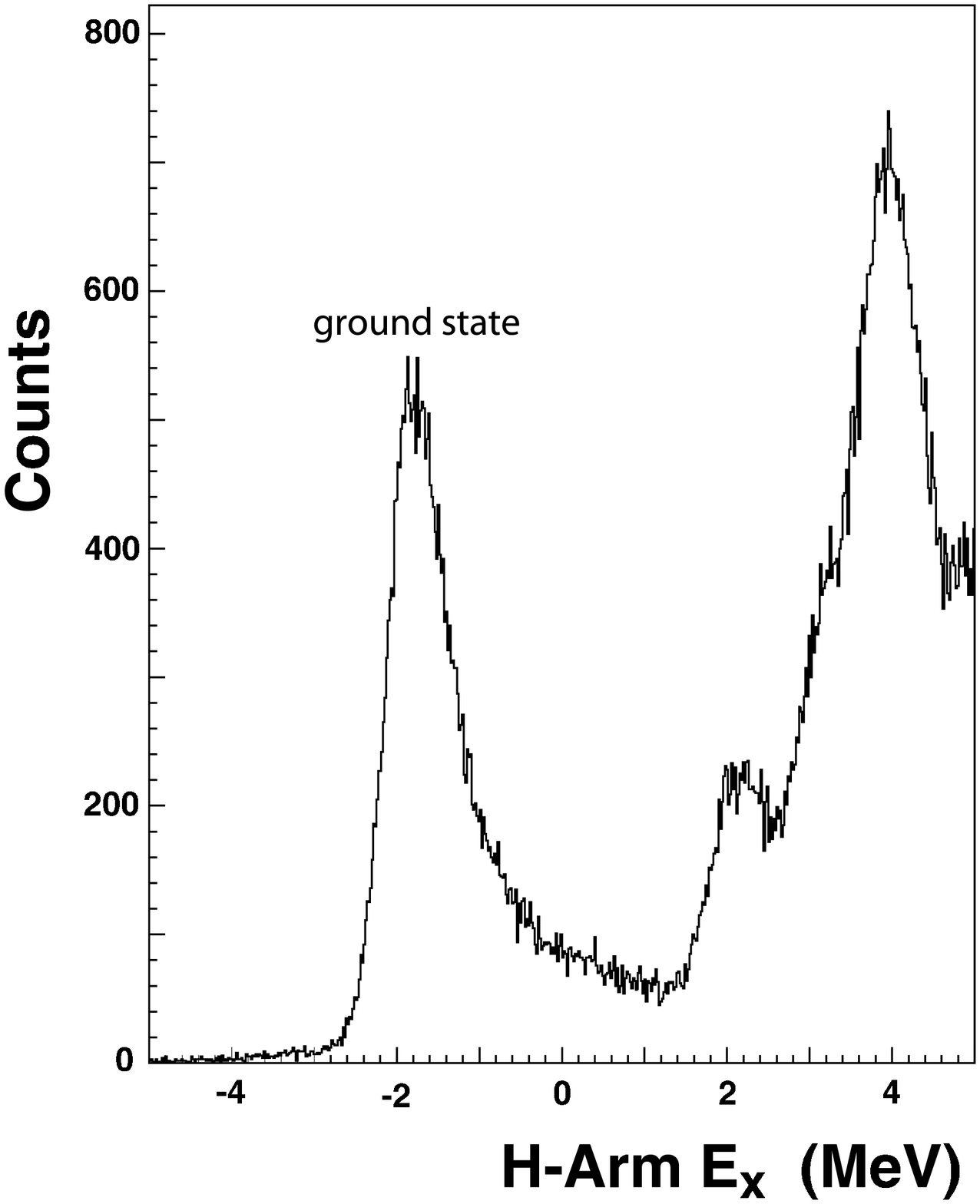}
\end{center}
\begin{center}
{\bf $\bf ^3$He}
\end{center}
\begin{center}
\includegraphics[width=0.4\textwidth]{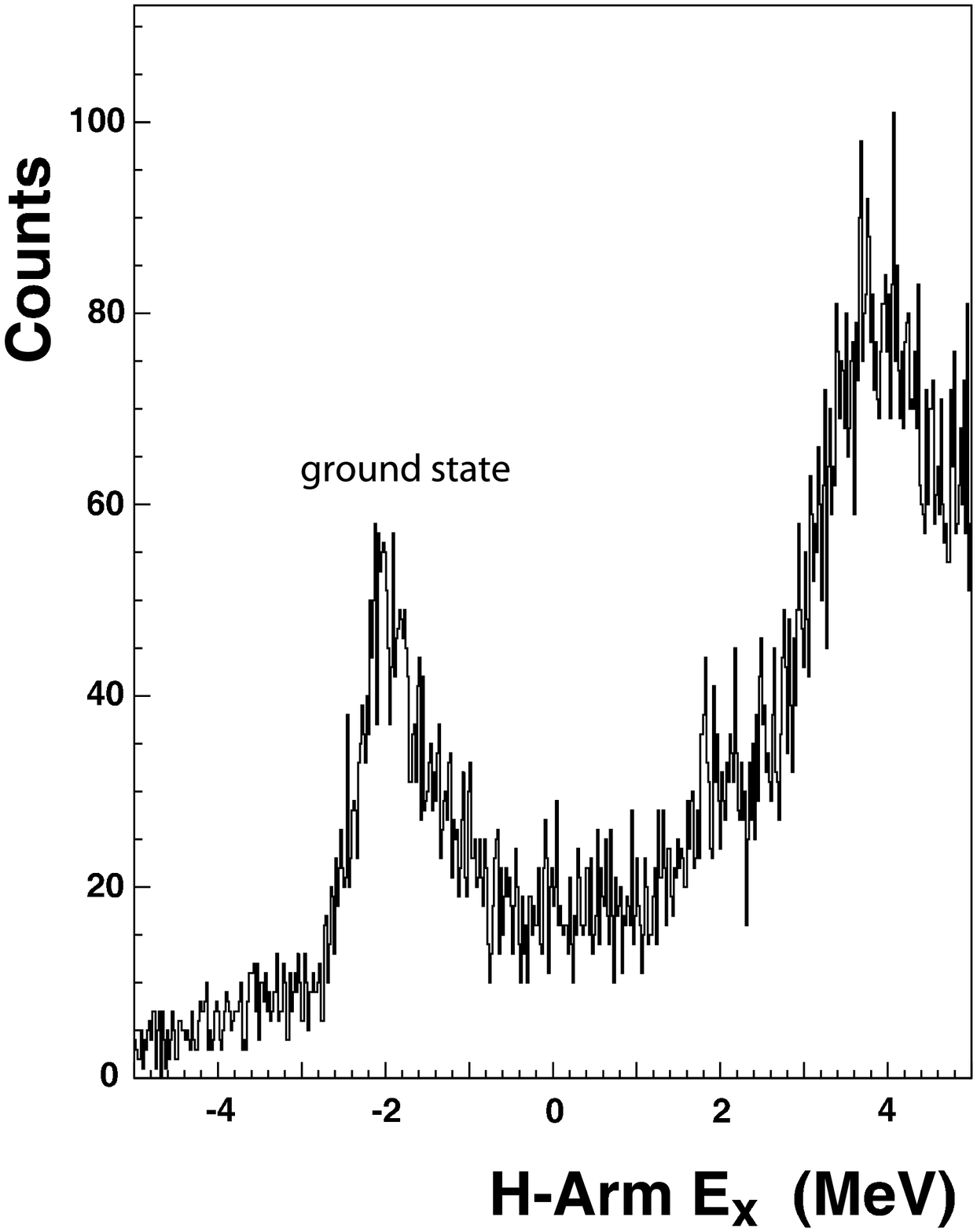}
\end{center}
\caption{\small{Raw yields measured with the right-arm spectrometer
in the region of the N$_2$ elastic peak 
using the N$_2$ reference cell (upper panel) and the 
$^3$He target (lower panel) as a function of excitation energy, $E_x$.
The leftmost peak represents the N$_2$ ground
state, and the other peaks are related to excited states of N$_2$}.}
\label{fig:n2he3}
\end{figure}

The nitrogen partial pressure in the $^3$He target cell
was determined as follows: As shown in Figure~\ref{fig:n2he3}, 
the elastic nitrogen peak was clearly resolved in both the
reference cell nitrogen spectrum (upper panel) and the spectrum
from the $^3$He target cell (lower panel), as measured with the
right-arm spectrometer.
As the nitrogen pressure corresponding to the reference cell spectrum was
known, the nitrogen pressure in the target cell could be determined
by simple scaling.  
This procedure was only required for one kinematic setting since the
nitrogen pressure was essentially constant throughout the experiment.
The result was $p_{N_2} = 15.15 \pm 0.35$~kPa.
The variation of the N$_2$ yield as a function of time was found to be 
within $\pm 3$\%. We assigned an overall uncertainty of 5\% to
the measured nitrogen background yields.

Obtaining the empty target yield ({\it i.e.}\/ the yield due to
scattering from the $^3$He target cell walls) 
from the empty reference cell data was complicated by two factors: 
(1) the background yield from the cell walls was a function
of beam position and the beam tune, and thus reference cell runs
did not necessarily reflect the exact background conditions 
present during production data taking; and
(2) the reference cell glass wall thickness and density were not equal to
those of the target cell.

Regarding (1), the variation of the empty
target yields obtained under nominally identical experimental
conditions but at different points in time were compared and found to
agree within $\pm 15$\%.

Regarding (2) we note 
(a) the target and reference cells were made of different types of
glass, where the target cell glass density was about 9\% larger than
that of the reference cell;
(b)  the thickness of the reference cell side walls was found to be,
on average, 2.5\% thinner than that of the target cell, as determined by laser 
interferometry \cite{xu_thesis}; and 
(c) the target cell had very thin (135~$\mu$m) end windows, while
the corresponding reference cell end windows 
were about as thick (1.2~mm) as its side walls.  
End window contributions were minimized by using software cuts.  The residual
contribution of the thick end windows tends to compensate the effect of the
thinner and less dense glass walls of the reference cell, although this
is difficult to quantify.  Hence,
we assumed that the empty cell background yield of the $^3$He target
was identical to that of the empty reference cell (without the need
for an explicit correction for the different cell properties)
and assumed an overall systematic uncertainty
of 25\% in the empty yield, taking into consideration the statistical 
uncertainty, the time variation of the yield due to beam 
tune variations, and the differences in the cell properties.

The empty target cell and the N$_2$ dilution factors ($R^{emp}$ and
$R^{N_2}$) were determined by combining all empty target and nitrogen
runs, respectively, at the same kinematics. The nominator in Equations
(\ref{eq:remp}) and (\ref{eq:rn2}) was calculated 
according to (\ref{eq:yields})
as $N^{qe}+N^{ert} = N - N^{emp} - N^{N_2}$.
The time variation
of the yields was included in the systematic uncertainty of each contribution.
The uncertainties are given in Table~\ref{tab:Aterror}.

An ad-hoc upward correction of all the empty target dilution factors by a
factor of 2, which was used in our prior publications
\cite{xu_thesis,Xu2000,Xu2003}, was dropped in this analysis as it had been
motivated by an unphysical tail of apparently poorly reconstructed events seen
in the right-arm spectrometer. Instead, a more conservative 
uncertainty  was assigned to the empty target background subtraction
at $Q^2 = 0.1$ and $0.2$~(GeV/c)$^2$, where the empty target 
background is largest.

\subsection{Monte Carlo Simulation}
\label{sec:montecarlo}
A full Monte Carlo simulation was developed for this experiment \cite{xiong},
which allowed averaging theoretical results over the experimental acceptances 
and accounted for multiple scattering, ionization energy loss, 
external bremsstrahlung, and internal radiative corrections.

To calculate the spin-dependent elastic and quasi-elastic 
radiative tails, internal radiation effects were modeled
using the covariant formalism developed in \cite{aku1}, 
generalized to the case of low-$Q^2$ quasi-elastic scattering.
This formalism accommodates polarization degrees of freedom.
Standard, unpolarized radiative corrections \cite{schwinger1}
were applied to the elastic peak region.

\subsection{Elastic Polarimetry} 
\label{sec:elastic_polarimetry}

\begin{table*}
\begin{center}
\begin{tabular}{|c|c|c|c|c|c|}
\hline\hline
$Q^2_{qe}$ & $Q^2_{el}$ & 
   \multicolumn{4}{c|}{$|A_{el}^{exp}(\%)|$} \\
(GeV/c)$^2$ & (GeV/c)$^2$ & 
   $-62.5^\circ$ / in  & $-62.5^\circ$ / out & 
   $-243.6^\circ$ / in & $-243.6^\circ$ / out \\
\hline
0.1 &  0.1 & $1.333\pm0.027$ & $1.043\pm0.027$ & $1.067\pm0.02$ & $1.208\pm0.030$ \\
0.193 &0.1 & $1.078\pm0.037$&$1.177\pm0.027$&$1.190\pm0.021$&$1.102\pm0.023$\\
0.3 &  0.2 & $1.251\pm0.096$ & $1.222\pm0.048$ & $1.107\pm 0.067$ & $1.206\pm0.075$\\
0.4 &  0.2 & $1.181\pm0.055$ & $1.314\pm0.061$& $1.168\pm0.06$ & $1.258\pm0.057$\\
0.5 &  0.2 & $1.265\pm0.042$&$ 1.307\pm0.039$ & $1.200\pm0.045$& $1.184\pm0.041$\\
0.6 &  0.2 & $1.258\pm0.049$&$1.301\pm0.047$&$1.110\pm0.05$&$1.096\pm0.05$\\
\hline\hline
\end{tabular}
\end{center}
\caption{\small The measured elastic asymmetries $|A_{el}^{exp}|$ for the
six quasi-elastic kinematic settings.
$Q^2_{qe}$ and $Q^2_{el}$ are the momentum transfers
of the quasi-elastic and elastic measurements, respectively. 
The four columns of results
correspond to the four combinations of the signs of the
target spin and beam helicity. The column headings indicate the 
laboratory target spin angle and the position of the accelerator 
injector half-wave plate. The uncertainties are statistical.}
\label{tab:elasresult}
\end{table*}

The beam and target polarizations, $P_b$ and $P_t$, were monitored 
continuously during the experiment using elastic polarimetry.
As the $^3$He elastic form factors, 
the charge form factor $F_c$ and the magnetic form factor $F_m$, are known 
very well experimentally~\cite{amroun2}, the $^3$He elastic 
asymmetry can be calculated as \cite{donnelly1}
\begin{widetext}
\begin{equation}
A_{el} = \frac{-2\tau v_{T'} \cos\theta^\ast 
  \mu_A^2 F_m^2 
  +2\sqrt{2\tau(1+\tau)} v_{TL'}\sin\theta^\ast \cos\phi^\ast \mu_A Z 
  F_m F_c}
{(1+\tau) v_L Z^2 F_c^2 + 2\tau v_T \mu_A^2 
  F_m^2}.
\label{eq:elasym}
\end{equation}
\end{widetext}
Here, the $v_i$ are kinematic factors, $\tau = Q^2/4M^2_{^3{\rm He}}$, and
$\mu_A = \mu_{^3{\rm He}}(M_{^3{\rm He}}/M_N) = -6.37$.
To allow direct comparison with data, the 
Monte Carlo program described in Section~\ref{sec:montecarlo}
was used to average Equation~(\ref{eq:elasym})
over the experimental acceptance.
We then obtained
\begin{eqnarray}
P_bP_t = \frac{A_{el}^{exp}}{A_{el}^{sim}}\times f_{N_2}f_{emp},
\label{eq:pbpteq}
\end{eqnarray}
where $A_{el}^{exp}$ and $A_{el}^{sim}$ are the measured and
simulated elastic asymmetry, respectively,
and $f_{N_2}$ and $f_{emp}$ are correction factors for 
the measured nitrogen and empty target cell dilution, respectively, 
for the elastic data sets.

The data for $A^{exp}_{el}$ are listed in Table~\ref{tab:elasresult}. 
Separate data are shown for each of the four possible spin and helicity
configurations, which are largely consistent
within their errors. For the evaluation of
(\ref{eq:pbpteq}), the weighted average of the data for the four
spin combinations was used.
The dilution factors $f_{N2}$ and $f_{empty}$ were obtained
using the procedure described in Section~\ref{sec:dilution}. 

No radiative corrections were applied to the elastic data since most radiative
effects were included in the simulation. Missing
is the spin dependence of the Schwinger correction, which we
deemed negligible. 

At the two beam energies, $E=0.778$ and $E=1.727$ GeV, the overall relative
systematic uncertainty in $P_bP_t$ was 1.3\% and 1.7\%, respectively. In each
case, the dominant contribution came from the uncertainty in the form factors
$F_c$ and $F_m$, followed by the contribution from the uncertainty in the 
target spin direction.

The average $P_bP_t$ so obtained was $0.208 \pm 0.001 \pm 0.004$,
where the errors are statistical and systematic, respectively.
As a cross check, independent measurements of the polarizations were obtained
using M{\o}ller beam polarimetry and NMR target polarimetry, yielding
an overall average value of $P_bP_t = 0.215 \pm 0.013$ \cite{Ozkul2000}.
The elastic polarimetry results were used for further analysis and
averaged for each quasi-elastic kinematic
setting separately ({\it cf.}\/ Table~\ref{tab:elasresult}) to account for
possible slow changes of the polarizations with time. The observed stability of
the polarization data suggests that this procedure was adequate.

\section{Asymmetry Results}

\begin{figure*}
\begin{center}
\includegraphics[width=0.7\textwidth]{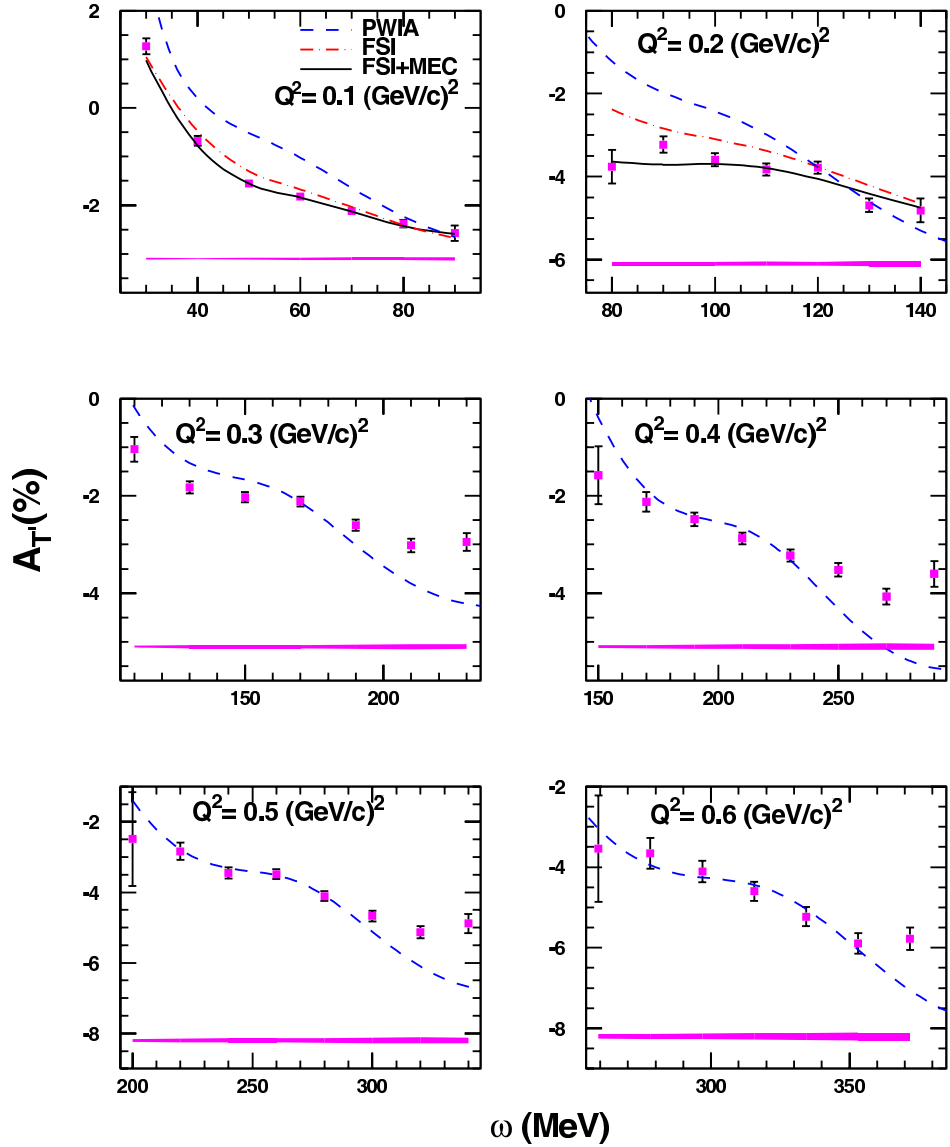}
\end{center}
\caption{\small{(Color online.) 
Quasi-elastic $A_{T'}$ asymmetry results vs.\ the energy 
transfer $\omega$. Errors on the data points are statistical. The 
systematic uncertainty is shown as an error band at the bottom of each 
panel.}}
\label{fig:At}
\end{figure*}

\subsection{Quasi-elastic Transverse Asymmetry A$_{T'}$}
Results for the quasi-elastic transverse asymmetry $A_{T'}$ at the six 
measured $Q^2$-points are shown in Figure~\ref{fig:At}.
Numerical values can be found in \cite{xu_thesis}.
The errors on the data are statistical only, while
the systematic uncertainty is shown as an error band at the bottom of 
each panel. A detailed breakdown of the systematic uncertainties
is presented in Table~\ref{tab:Aterror}. The experimental data
were corrected for radiative effects, background, and dilution, as
described in detail in the previous section.

Also shown in Figure~\ref{fig:At} are the results
of several calculations. Dashed lines represent the PWIA calculation
\cite{Kievsky97}.
The dash-dotted and solid curves at the two kinematics with lowest $Q^2$
represent, respectively,
Faddeev results with inclusion FSI only \cite{golak2} and with 
inclusion of both FSI and MEC corrections \cite{Golak01}.
Calculation \cite{Golak01}
will be referred to as the ``full Faddeev calculation'' in the following.
All theory results were averaged over the spectrometer acceptances
using the Monte Carlo simulation described in Section~\ref{sec:montecarlo}.
Further details on the calculations are given in Section~\ref{sec:theory}.

One observes excellent agreement of the data with the full 
Faddeev calculation over the entire $\omega$-range 
at $Q^2 = 0.1$ and $0.2$ (GeV/c)$^2$, 
while PWIA describes the data well at the higher $Q^2$, in
particular in the region around the quasi-elastic peak 
(near the center of the $\omega$-range in each panel).

\subsection{Asymmetry in the Threshold Region}
\label{sec:thresholdasym}

\begin{figure*}
\centerline{\includegraphics[width=0.53\textwidth,angle=0]{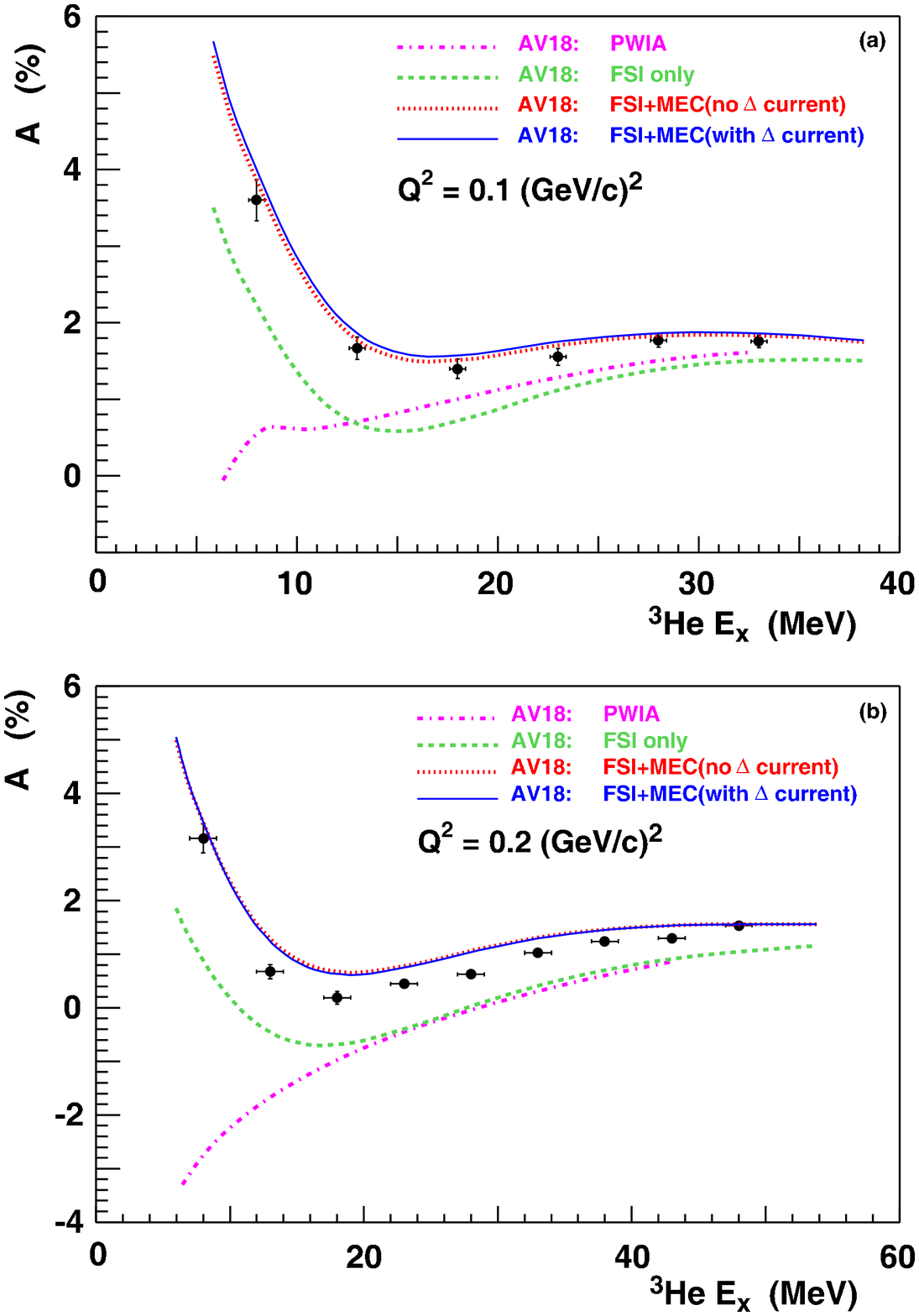}}
\caption{(Color online.) 
The experimental asymmetry in the region of the $^3$He breakup
 threshold together with theoretical calculations for 
     (a) $Q^2$ = 0.1 (GeV/c)$^2$
 and (b) $Q^2$ = 0.2 (GeV/c)$^2$. 
The calculations differ only in the description of the reaction mechanism.
} 
\label{fig:threshasy}
\end{figure*}

The asymmetries measured in the region around the
two- and three-body breakup thresholds (5.5 and 7.7 MeV, respectively)
are shown in Figure~\ref{fig:threshasy}. These results 
provide a sensitive test of the quality of the Faddeev calculations.

The threshold asymmetry data were taken with the hadron-arm spectrometer
as a by-product of the elastic polarimetry and were analyzed in the
same manner as the quasi-elastic asymmetries.
The kinematics are given in the lower panel of
Table~\ref{tab:kinematics}. The $Q^2$-values of 0.1 and 0.2 (GeV/c)$^2$
in Figure~\ref{fig:threshasy}
correspond to the momentum transfer at the elastic peak.
The data are plotted as a function of the excitation energy, $E_x$,
defined in Equation~(\ref{eq:ex}).
Horizontal errors represent the uncertainty in 
determining $E_x$, which was dominated by the uncertainty in the beam 
energy. The vertical errors are the statistical and systematic errors added in 
quadrature. Tables of the data and uncertainties can be found 
in \cite{Feng2002}.

Figure~\ref{fig:threshasy} also shows various theoretical results.
Dot-dashed lines depict those of the PWIA calculations~\cite{Kievsky97},
while those of Faddeev calculations with FSI only~\cite{golak2} appear
as dashed lines. The ``full Faddeev calculation'', which includes both FSI 
and MEC, but not the $\Delta$-isobar current, \cite{Golak01} 
is represented by dotted lines.
The solid lines were obtained with the full calculation after 
including $\Delta$-isobar currents.
These calculations employed the AV18 NN interaction 
potential. Results obtained with the BonnB potential
were found to be only slightly different
from the AV18 results and in even better agreement with the data \cite{xiong}.

As can be seen, the agreement between PWIA calculations
and the data is poor at both kinematics, which confirms the expectation that
FSI and MEC corrections are essential in this region. 
Indeed, the inclusion of FSI improves the 
agreement significantly, and good agreement is achieved once MEC are added.
It has been shown that substantial 
MEC are needed to describe the measured elastic 
electromagnetic form factors of three-nucleon systems \cite{amroun1}.
The corresponding physics should extend into the low-$\omega$
region of inelastic scattering as well.

The good agreement between the full calculation and the data at $Q^2$ = 
0.1 (GeV/c)$^2$ suggests that FSI and MEC are properly treated in the 
full calculation. The insensitivity of the results to the 
addition of $\Delta$-isobar currents implies a weak model dependence of 
the MEC corrections. The small, systematic discrepancy at 
$Q^2$ = 0.2 (GeV/c)$^2$ may indicate that some $Q^2$-dependent 
effects, such as relativistic and three-nucleon force effects, 
become important already at this momentum transfer. 

\section{Extraction of the Neutron Magnetic Form Factor}
\label{sec:ffextraction}

The neutron magnetic form factor, $G_M^n$, can be extracted from
the measured $^3$He quasi-elastic transverse 
asymmetry $A_{T'}$ if a calculation is available that
predicts $A_{T'}$ as a function of $G_M^n$. 
If we assume, following Equation~(\ref{eq:at_heuristic}), 
that the asymmetry is a function of $({G_M^n})^2$, 
we can expand $A_{T'}$ around a reference $G_M^n$
value, $G_0$,
\begin{eqnarray}
  A_{T'} ({G_M^n}^2) & = &   A_{T'} (G_0^2)
        +\frac{\partial A_{T'}}{\partial ({G_M^n}^2)} (G_0^2) \times
        ({G_M^n}^2-G_0^2) \nonumber\\
        &+&O ( ({G_M^n}^2-G_0^2)^2).
\label{eq:gmnexpand}
\end{eqnarray}
For ease of notation, we normalize
all $G_M^n$ values to a convenient reference scale
(the H\"{o}hler parameterization \cite{hoh} in this case) so that
$G_0 = 1$. Equation~(\ref{eq:gmnexpand}) 
can be solved for $G_M^n$, assuming the second-order term is small:
\begin{equation}
  G_M^n = \sqrt{1 + \frac{A_{T'}({G_M^n}^2) - A_{T'}(1)}
                      {\partial A_{T'}/\partial ({G_M^n}^2)(1)}}.
\label{eq:gmnsolv}
\end{equation}
Here, $A_{T'}({G_M^n}^2)$ is the measured asymmetry.
The predicted asymmetry, $A_{T'}(1)$, and the sensitivity 
factor, $\partial A_{T'}/\partial ({G_M^n}^2)(1)$, are the output of
the model calculation. 
The latter two parameters were determined using the full Faddeev 
calculation \cite{Golak01} for the lowest two $Q^2$ points
of this experiment, and the PWIA calculation \cite{Kievsky97}
for the remaining four $Q^2$.
Results were averaged over the experimental acceptance
using the Monte Carlo program described in Section~\ref{sec:montecarlo}.
At each kinematical point, asymmetries were generated for several 
$\omega$-bins around the quasi-elastic peak. Within each bin, $G_M^n$ 
was varied around the reference value $G_0$ by adding a constant to
the functional form $G_M^n(Q^2)$ given by the H\"{o}hler model.

$G_M^n(Q^2)$ was extracted for each $\omega$-bin via (\ref{eq:gmnsolv}).
A different functional form, a general second-order expansion of 
$A_{T'}(G_M^n)$, was also tried. The differences between the form factors 
extracted via these two methods was found to be negligible ($< 0.1$\%) 
for all kinematics \cite{xu_thesis}

The final $G_M^n$ results
were obtained by taking the weighted average of the $G_M^n$ values
from the $\omega$-bins closest to the quasi-elastic peak.
The $\omega$-region used
for the extraction of $G_M^n$ covered a width of 30~MeV at $Q^2 = 0.1$ 
and 0.2, 60~MeV at $Q^2 = 0.3$, 40~MeV at $Q^2 = 0.4$ and 0.5,
and 56.25~MeV at $Q^2 = 0.6$ (GeV/c)$^2$.

The extraction procedure gives rise to a
systematic error due to the uncertainty in the 
experimental determination of the energy transfer $\omega$ ($\pm 3$~MeV).
The uncertainty in $\omega$  results in an
uncertainty as to the $\omega$-region over which to integrate the
theoretical calculation used for the extraction of $G_M^n$.
A shift of bin boundaries generally
translates into a different average value of $A_{T'}$ for the bin and hence
a different extracted $G_M^n$ value. 

Furthermore, as can be seen in Figure~\ref{fig:At}, the theoretical 
calculations, especially PWIA,
match the data best in the immediate vicinity of the quasi-elastic
peak where corrections to the plane-wave picture are smallest,
whereas deviations may occur off the peak.
This can introduce an artificial $\omega$-dependence into the
extracted $G_M^n$ which goes beyond the effect of the 
kinematical variation of $Q^2$ with $\omega$. 
For this effect to be minimized, 
the bins used for the $G_M^n$ extraction should be centered 
around the quasi-elastic peak, assuming that
deviations are distributed roughly symmetrically.
The experimental uncertainty in $\omega$ may cause improper centering, 
resulting in a bias in extracting $G_M^n$.
The calculated uncertainties in $G_M^n$ resulting from the uncertainty
in $\omega$  can be found in Table~\ref{tab:gmn_uncertainties}.

\begin{table}
\begin{center}
\begin{tabular}{|ccc|}
\hline\hline
Source   &  \multicolumn{2}{c|}{$\delta A_{T'}/A_{T'} (\%)$}  \\
         & $Q^2 \leq 0.2$ \hspace{3mm}  & $Q^2 \geq 0.3$  \\
\hline
$P_{t}P_{b}$                    & 1.3  &  1.7              \\
Empty target subtraction        & 1.0  &  0.25             \\
N$_2$ background subtraction    & \multicolumn{2}{c|}{0.3} \\
QE radiative correction         & \multicolumn{2}{c|}{0.3} \\
Elastic radiative tail          & \multicolumn{2}{c|}{0.3} \\
Spectrometer acceptance         & \multicolumn{2}{c|}{0.5} \\
HC scintillator efficiency      & \multicolumn{2}{c|}{0.1} \\
HC wire chamber efficiency      & \multicolumn{2}{c|}{0.1} \\
HC computer deadtime            & \multicolumn{2}{c|}{0.1} \\
HC beam current shift           & \multicolumn{2}{c|}{0.1} \\
HC beam motion                  & \multicolumn{2}{c|}{0.1} \\ 
Pion contamination              & \multicolumn{2}{c|}{0.1} \\
\hline
Total                           & 1.8  &  1.9              \\
\hline
\hline
\end{tabular}
\end{center}
\caption{\small{Estimated systematic uncertainties 
of the quasi-elastic $A_{T'}$ asymmetry measurements. 
``HC'' denotes ``he\-li\-ci\-ty-correlated''. The two columns of uncertainties
correspond to the quasi-elastic measurements at lower 
$Q^2$ ($0.1$ and $0.2$) and higher $Q^2$ ($0.3 - 0.6$), respectively.
Values in the center of both columns are common to all kinematics.}}
\label{tab:Aterror}
\end{table}

\section{Estimate of Theoretical Uncertainties}

\subsection{Nucleon-Nucleon Potential and Nucleon Form Factors}

The effect of different NN potential models on the
predicted asymmetry $A_{T'}$ was studied by carrying out the
full Faddeev calculation with the Argonne AV18 and the
Bonn B NN potentials at several representative kinematics.
In a similar manner, to estimate the uncertainty due to the 
elastic nucleon form factors other than $G_M^n$, 
Faddeev calculations were performed in which these quantities were
varied individually by their published experimental uncertainties.
The resulting uncertainty in $G_M^n$ from these sources,
when combined in quadrature, is less than 1\% for all kinematics
(cf.\ Table~\ref{tab:gmn_uncertainties}).

\subsection{Relativistic Effects}
\label{sec:rel-effects}

Since the full Faddeev calculation
is non-relativistic, it was particularly important to estimate quantitatively
the size of relativistic corrections.
An approximate estimate can be obtained within 
PWIA, which is theoretically well understood. Standard PWIA calculations
take most relativistic effects into account ({\it cf.}\/ 
Section~\ref{sec:pwia}).
It is straightforward to modify the relativistic parts of the PWIA 
formalism to reflect the non-relativistic approximations
made in the Faddeev formalism. The differences between the results
of such a modified, non-relativistic PWIA calculation and the
standard relativistic PWIA results provide an estimate of the
error in the Faddeev results due to relativistic effects.

\begin{figure}
\begin{center}
\includegraphics[width=0.48\textwidth]{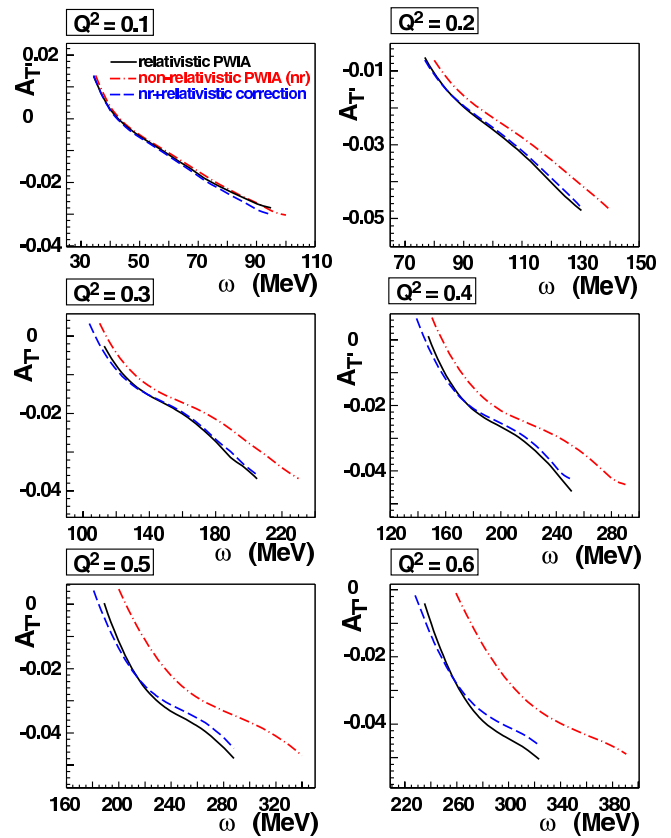}
\end{center}
\caption{\small (Color online.) Relativistic effects in $A_{T'}$. The solid 
line is the standard, relativistic PWIA calculation \protect\cite{Kievsky97}. 
The dot-dashed curve is the non-relativistic PWIA calculation that 
we developed, and the dashed curve is the non-relativistic 
PWIA calculation with heuristic relativistic corrections applied (see text).}
\label{fig:relasy}
\end{figure} 
          
To this end, we modified three parts of the standard PWIA formalism:
approximations 
were made to the relativistic kinematics, the phase space 
and the integral ranges of the Fermi momentum and the missing mass of
the many-fold integration of the $^3\vec{\rm He} (\vec{e},e')$
cross-section were changed according to the non-relativistic
kinematics, and the relativistic hadronic current was translated into
an approximate, non-relativistic form \cite{Jeschonnek98,Ritz97}.
Among the three modifications, the change of the
kinematics was found to dominate \cite{Donnelly}.

With the PWIA results at hand,
we developed a heuristic ``recipe'' \cite{Donnelly} to allow an approximate
correction of the Faddeev results for relativistic effects. The ``recipe''
could be readily applied to existing Faddeev results 
without the need for recomputation.

Results of these studies are shown in Figure~\ref{fig:relasy}.
The three curves represent the original relativistic PWIA results
(solid line), non-relativistic PWIA results obtained using the
modifications described above (dot-dashed line),
and non-relativistic PWIA results corrected for relativistic effects
though the ``recipe'' (dashed line).
As can be seen, the heuristic correction works well
up to about $Q^2 = 0.4$ (GeV/c)$^2$.

\begin{figure}
\begin{center}
\includegraphics[height=0.45\textheight]{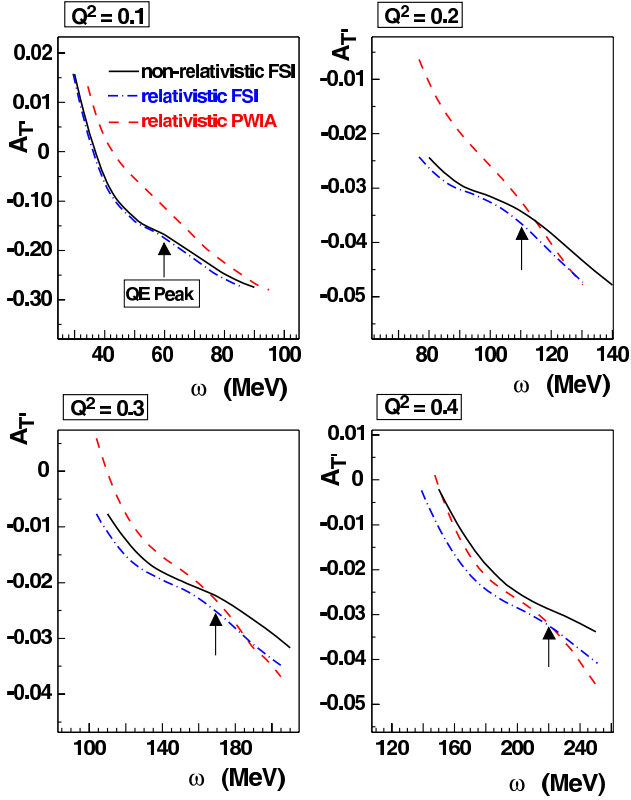}  
\end{center}
\caption{\small (Color online.) FSI effect study. The dashed curve 
represents the standard (relativistic) PWIA calculation, 
the solid curve is the (non-relativistic) Faddeev
calculation with FSI effects only, and the dot-dashed curve depicts 
the same Faddeev calculation but with relativistic
corrections applied. Comparing the dashed and dot-dashed curves, one can
estimate the effect of FSI in $A_{T'}$.}
\label{fig:fsi}
\end{figure}    

The acceptance-averaged difference between the relativistic and 
non-re\-la\-ti\-vis\-tic PWIA results
at $Q^2 = 0.1$ and 0.2 (GeV/c)$^2$
was taken as the model uncertainty of the Faddeev
results due to relativity.

\subsection{FSI \& MEC}

\begin{figure}
\begin{center}
\includegraphics[width=0.4\textwidth]{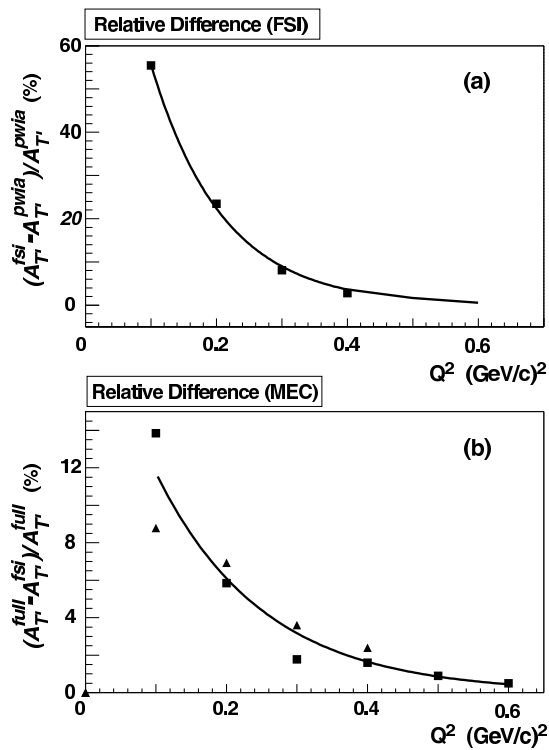}
\end{center}
\caption{\small{Estimated magnitude of (a) FSI effects
and (b) MEC effects in $A_{T'}$ as a function of $Q^2$.
In the lower panel (b), the solid triangles represent results
obtained in our study of $^3\vec{\rm He}(\vec{e}, e')$,
while the solid squares depict predictions from a 
calculation of $\vec{d} (\vec{e}, e')$ obtained in 
\protect\cite{Arenhovel93}. The curves are empirical fits to the data.}}
\label{fig:fsiex}
\end{figure}

To estimate FSI contributions to $A_{T'}$, we carried out
the Faddeev calculation up to $Q^2 = 0.4$ (GeV/c)$^2$ with the 
inclusion of FSI effects only.
(Already at $Q^2 = 0.3$~(GeV/c)$^2$, the 3N center-of-mass energy is above
the pion production threshold, and therefore the non-relativistic
framework is no longer valid.)
Next, we applied relativistic corrections to the Faddeev results using
the ad-hoc prescription developed in Section 
\ref{sec:rel-effects}. The ``relativistic FSI'' results so obtained
were compared to the results of the standard (relativistic) PWIA
calculation, as illustrated in Figure~\ref{fig:fsi}. The 
difference between the two calculations in the region around the 
quasi-elastic peak is a measure of the FSI effects at each $Q^2$ point.
For the two highest $Q^2$ values, we extrapolated the FSI data
using a purely empirical fit to the lower $Q^2$ values, as
shown in Figure~\ref{fig:fsiex}(a). As expected \cite{JRA,Pace91,Benhar99}, 
FSI effects decrease significantly as $Q^2$ increases.

\begin{figure}
\begin{center}
\includegraphics[height=0.45\textheight]{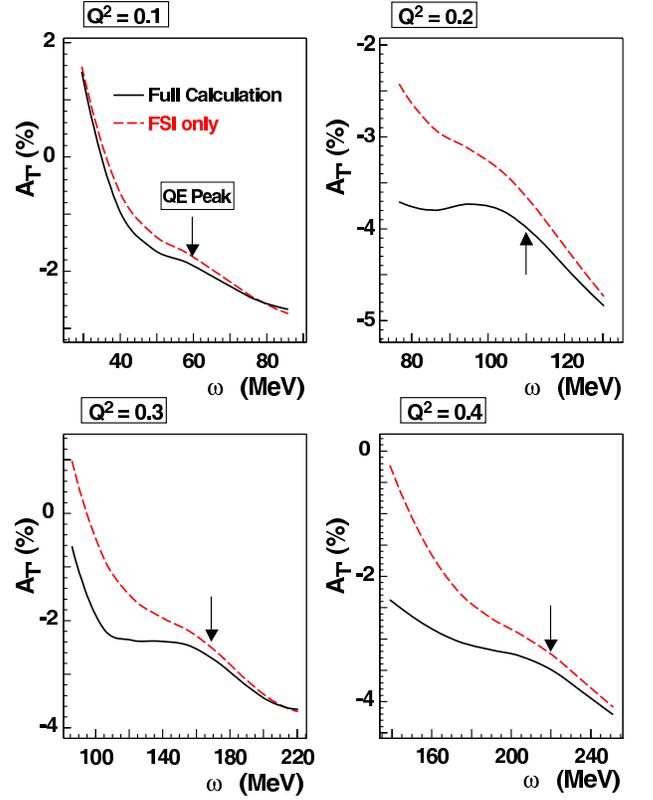}
\end{center}
\caption{\small{(Color online.) 
MEC effect study. Comparing the full calculation 
(solid curve) with the calculation with FSI effects only (dashed curve),
one can estimate contributions to $A_{T'}$ from MEC effects.}}
\label{fig:golakmec}
\end{figure}

In a similar manner, we can estimate the size of MEC effects by
comparing the Faddeev results with inclusion of FSI only, obtained in the
FSI study above, to those of the full Faddeev calculation.
Results are shown in Figure~\ref{fig:golakmec}, and differences 
between the two calculations are plotted as solid triangles 
in Figure~\ref{fig:fsiex}(b). As with FSI, we observe a
sharp decrease of MEC corrections with increasing $Q^2$.

It is interesting to compare our results for the size of MEC corrections
with those obtained from theoretical studies 
of quasi-elastic inclusive scattering from polarized deuterium, 
$\vec{d} (\vec{e}, e')$ \cite{Arenhovel93}. The deuterium results are shown in 
Figure~\ref{fig:fsiex}(b) as solid squares. 
As can be seen, the data are similar to those 
calculated for the corresponding $^3{\rm He}$ reaction.
Assuming a similar underlying physical mechanism,
we use the MEC data from deuterium to estimate the size of MEC corrections
to the $^{3}\vec{\rm He}(\vec{e}, e')$ data at the highest two $Q^2$
values of our data set.

These studies provide rough estimates of
the expected magnitudes of the respective effects. They are not
reliable enough to be used to correct the PWIA results
for FSI and MEC contributions. Consequently, we use the numbers 
obtained above as estimates of the model uncertainties inherent 
in the PWIA.  We take the numbers
as the 1$\sigma$ values of the uncertainties, which we assume to be
symmetric. The resulting model uncertainties in $G_M^n$
are detailed in Table~\ref{tab:gmn_uncertainties} and are
propagated into the final $G_M^n$ errors given in Table~\ref{tab:gmnresults}.

\begin{figure*}
\begin{center}
\includegraphics[height=0.6\textwidth, angle=90]{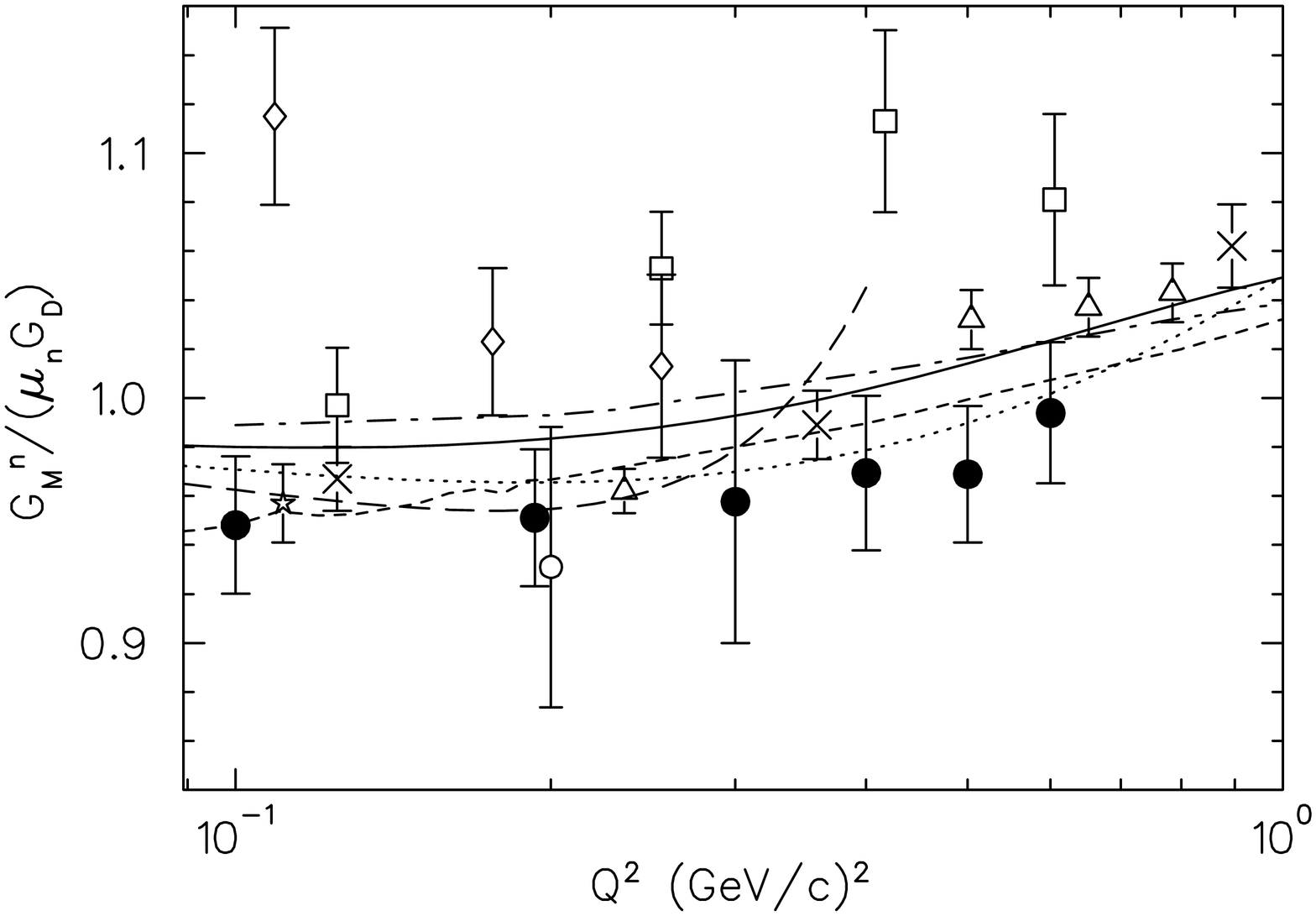}
\end{center}
\caption{\small{The world's $G_M^n$ data since 1990.
Data points represent the results of the Bonn \cite{Brui95,Schoch2000} 
($\Box$), MIT-Bates \cite{Mark93,Gao94} 
($\Diamond$, \protect\raisebox{0.7ex}{\protect\circle{5}}\!),
NIKHEF/PSI \cite{Ankl94} (\mbox{\ding{73}}) 
and the Mainz/PSI \cite{Ankl98,Kubon2002} ($\vartriangle$,$\times$) 
experiments as well as those of the present measurement (\mbox{\ding{108}}),
where the error bars are the total uncertainties reported.
Also shown are the results of various model calculations:
Hammer and Mei{\ss}ner \cite{Hammer_Dispersion} (solid curve),
Holzwarth \cite{Holzwarth} (dotted),
Kubis and Mei{\ss}ner \cite{Kubis2001} (long-dashed),
Lomon \cite{Lomon2002} (dashed-dotted), and
de Melo {\it et al.}~\cite{Salme2006} (short-dashed).}}
\label{fig:gmndata}
\end{figure*}

\begin{table}[b]
\begin{center}
\begin{tabular}{|c|c|c|c|c|c|}\hline\hline
$Q^{2}$& $G_M^n / (\mu_n G_D) $ &\multicolumn{4}{c|}{$\delta G_M^n/G_M^n$} \\
& & stat. & syst. & model & total \\ 
(GeV/c)$^2$ &   & (\%) & (\%) & (\%) & (\%) \\
\hline
0.1 & 0.9481 &  1.36 & 1.08 & 2.2 & 2.8 \\
\hline
0.193 & 0.9511  &  1.35 & 1.26 & 2.1 & 2.8 \\
\hline
0.3&   0.9577 & 1.35 & 1.86 & 5.3 & 5.8 \\
\hline
0.4&   0.9694 & 1.45  & 1.28 & 2.5 & 3.2 \\
\hline
0.5&   0.9689 & 1.35  & 1.25 & 2.1 & 2.8 \\
\hline
0.6&   0.9939 & 1.55   & 1.38 & 2.0 & 2.9 \\  
\hline\hline
\end{tabular}
\end{center}
\caption{\small{Results for $G_M^n$ as a ratio to the
dipole form factor, $G_D$, and uncertainties 
obtained in the present experiment. The data 
have changed slightly from our previously published numbers
\protect\cite{Xu2000,Xu2003} due to differences in the analysis.}}
\label{tab:gmnresults}
\end{table}                                        

\subsection{Off-Shell Effects}

Off-shell corrections to the single nucleon current,
including the part of the current that describes polarization degrees of 
freedom \cite{Caballero93}, are purely relativistic in nature.
While the PWIA calculation used here includes off-shell effects, 
they are ignored in the Faddeev calculation.

We estimated the magnitude of required off-shell corrections to the
Faddeev results by comparing results of a modified version of the 
PWIA formalism that treats nucleons as on-shell \cite{Donnelly} to
those of the standard PWIA.

In addition, theoretical uncertainties due to 
different possible off-shell prescriptions were estimated using the
difference of PWIA results obtained with the deForest CC1 and CC2 
forms \cite{deForest83}. (The standard PWIA calculation employs CC1.)
While this number represents a minimum
uncertainty, as various other off-shell prescriptions are equally
permissible \cite{Caballero93}, PWIA calculations using the
CC1 form have been found to agree better with experimental
data of unpolarized $^3{\rm He}(e,e')$ scattering than those using
other prescriptions \cite{Kievsky97}. This suggests the use of the CC1
prescription as a reference in the polarized case as well.

Results are given in Table~\ref{tab:gmn_uncertainties}.
Interestingly, off-shell effects dominate the model 
uncertainty in $G_M^n$ at the lowest two $Q^2$-values.

\begin{table*}
\begin{center}
\begin{tabular}{|c|cccccc|cccccccc|}
\hline\hline
$Q^2$       & \multicolumn{6}{c|}{systematic $\delta G_M^n/G_M^n$ (\%)}
            & \multicolumn{8}{c|}{model $\delta G_M^n/G_M^n$ (\%)} \\
(GeV/c)$^2$ & $A_{T'}$ & $\omega$ & $G_E^p$ & $G_M^p$ & $G_E^n$ & Total
            & {\it NN} & off-shell & FSI & MEC & 3BF 
            & Coulomb & Relativity & Total \\
\hline
0.1   &  0.90  & 0.3  & 0.44  & 0.21  & 0.14 & 1.08
      &  0.45  & 1.6 & 0.5 & 1.0 & 0.6
      &  1.0 & 0.5 & 2.2 \\
0.193 &  0.90  & 0.6  & 0.53 & 0.35 & 0.13 & 1.26
      &  0.40  & 1.2 & 0.5 & 1.0 & 1.0
      &  1.0  &  0.7  & 2.1 \\
0.3   &  0.95 &  1.4  & 0.56  &  0.52  &  0.17 & 1.86
      &  0.50 & 0.5  & 4.5  & 1.8 & 1.2
      &  1.0  & 0.5  & 5.3 \\
0.4   &  0.95  & 0.45  &  0.46 & 0.56  & 0.08 & 1.28
      &  0.45 & 0.5 &  1.8 & 1.2 & 1.2 
      &  1.0 & 0.5 & 2.5 \\
0.5   &  0.95  & 0.15  &  0.38 & 0.60  & 0.38 & 1.25
      &  0.40 & 0.5 & 0.7  & 0.5 & 1.4
      &  1.0 & 0.5 & 2.1 \\
0.6   &  0.95  & 0.10  &  0.32 & 0.64  & 0.69 & 1.38 
      &  0.40 & 0.5 & 0.5 & 0.5 & 1.4
      &  1.0 & 0.5 & 2.0 \\
\hline
\end{tabular}
\end{center}
\caption{\small Estimated uncertainties of the extracted
form factor $G_M^n$. 
Systematic uncertainties include contributions from
the asymmetry measurement ($A_{T'}$; see table \protect\ref{tab:Aterror}), 
the energy transfer determination ($\omega$), and
the other nucleon form factors ($G_E^p$, $G_M^p$, and $G_E^n$).
Theoretical (model) uncertainties include contributions from 
the {\em NN} potential model, off-shell effects, final-state interactions
(FSI), meson-exchange currents (MEC), three-body forces (3BF), Coulomb
corrections, and relativistic effects.
In the totals, the uncertainties have been added in quadrature, 
ignoring any possible
correlations between the contributions, which may very well exist,
especially for the model uncertainties. Thus, the numbers
should be taken with appropriate caution.}
\label{tab:gmn_uncertainties}
\end{table*}

\section{Form Factor Results and Discussion}

Numerical values for $G_M^n$ extracted in this work are given in 
Table~\ref{tab:gmnresults} 
(in units of the empirical dipole parameterization, 
$G_D = (1+Q^2/0.71)^{-2}$)
and shown in Figure~\ref{fig:gmndata}
along with the existing world data set published since 1990
\cite{Mark93,Ankl94,Gao94,Brui95,Schoch2000,Ankl98,Kubon2002}.
The error bars 
represent the total uncertainties reported by the respective experiments,
including model uncertainties.

The results appear to be largely consistent, with
the exception of the early $^2$H$(e,e'n)$ data from Bates \cite{Mark93}
and the first $^2$H$(e,e'n)/^2$H$(e,e'p)$ ratio measurement from Bonn 
\cite{Brui95,Schoch2000}.
The discrepancy of the data of these two experiments with the rest
of the world data has been attributed to incomplete
corrections for neutrons that miss the neutron detector \cite{jourdan}.
The data of the Bonn experiment \cite{Brui95} 
were re-analyzed subsequently \cite{Schoch2000},
resulting in a downward correction of the $G_M^n$ data. 
Figure~\ref{fig:gmndata} shows the re-analyzed data.
We note the satisfactory agreement between
the more recent, high-precision deuterium ratio measurements 
\cite{Ankl94,Ankl98,Kubon2002} and the data from this work.
The agreement is well within the total uncertainties of the experiments,
except at $Q^2 = 0.5$ and $0.6$~(GeV/c)$^2$, where the $^3$He results
are low by about 5\%.

Also shown in Figure~\ref{fig:gmndata} are several theoretical results:
a recent dispersion-theoretical fit
by Hammer and Mei{\ss}ner \cite{Hammer_Dispersion} (solid curve),
a chiral soliton model by Holzwarth \cite{Holzwarth} (dotted
curve), a relativistic baryon chiral perturbation theory calculation by 
Kubis and Mei{\ss}ner \cite{Kubis2001} (long-dashed curve),
a vector meson dominance (VMD) fit by Lomon \cite{Lomon2002} (dashed-dotted
curve), and a recent light-front quark model by de Melo 
{\it et al.}~\cite{Salme2006} (short-dashed curve). 
It should be noted that all of these models contain one
or more free parameters that have been fitted to existing data.

As can be seen, the dispersion-theoretical fit \cite{Hammer_Dispersion}
and the VMD fit \cite{Lomon2002} agree best with the data
at $Q^2 > 0.3$ (GeV/c)$^2$, while the chiral perturbation theory (ChPT) 
results \cite{Kubis2001} and the chiral soliton model \cite{Holzwarth}
match the data better at lower $Q^2$.  
The ChPT model is expected to be good only up to $Q^2 \approx 0.3$~(GeV/c)$^2$, 
but clearly works very well in its region of validity. The light-front model of
de Melo {\it et al.}~\cite{Salme2006} arguably shows the best overall
agreement. The models by Holzwarth \cite{Holzwarth}, Lomon \cite{Lomon2002}, 
and de Melo {\it et al.}~\cite{Salme2006} also describe the proton form factor
ratio $G_E^p/G_M^p$ and other elastic nucleon form factors well in 
this $Q^2$-region.

\section{Conclusions}

In conclusion, we have determined the neutron magnetic form 
factor $G_M^n$ from quasi-elastic $^3\vec{\rm He}(\vec{e},e')$ data.
At $Q^2$ of 0.1 and 0.2 (GeV/c)$^2$, we used a 
state-of-the-art Faddeev calculation that includes FSI and MEC, 
and PWIA at four additional points between $Q^2 = 0.3$ and 0.6 (GeV/c)$^2$.
The results agree within the total uncertainties with those obtained by
several recent measurements on deuterium, except at $Q^2 = 0.5$ and $0.6$
(GeV/c)$^2$ where the $^3$He results are slightly low.  A consistent picture
of the behavior of $G_M^n$ in this $Q^2$-region is beginning to emerge,
although further precision measurements as well as improved model
calculations, such as the extension of the Faddeev formalism to higher $Q^2$
in the case of polarized $^3{\rm He}$, remain highly desirable.

In addition, we have measured $A_{T'}$ in the two- and three-body breakup
threshold region at $Q^2$ of 0.1 and 0.2 (GeV/c)$^2$ where the sensitivity
to FSI and MEC effects is particularly high.  The results 
agree well with the predictions of the Faddeev model, especially
at $Q^2 = 0.1$~(GeV/c)$^2$, confirming the validity of the treatment of
FSI and MEC effects in this formalism.

\begin{acknowledgments}

We thank the Hall A technical staff and the Jefferson Lab Accelerator
Division for their outstanding support during this experiment.  We
also thank T.~W.~Donnelly for many helpful discussions.  This work was
supported in part by the US Department of Energy, DOE/EPSCoR, the
US National Science Foundation, the Science and Technology
Cooperation Germany-Poland and the Polish Committee for Scientific
Research under grant no.\ 2P03B00825, 
the Ministero dell'Universit\`{a} e della Ricerca
Scientifica e Tecnologica (Murst), the French Commissariat \`{a}
l'\'{E}nergie Atomique, Centre National de la Recherche Scientifique
(CNRS), Conseil R\'egional d'Auvergne, the Italian Istituto Nazionale
di Fisica Nucleare (INFN), a grant of the European Foundation Project
INTAS-99-0125, and by DOE contract
DE-AC05-84ER40150, Modification No.\ M175, 
under which the Southeastern Universities Research
Association (SURA) operates the Thomas Jefferson National Accelerator
Facility.  The numerical calculations were performed on the PVP
machines at the US National Energy Research Scientific Computer
Center (NERSC) and the CRAY SV1 of the NIC in J\"{u}lich.

\end{acknowledgments}


\end{document}